\documentclass[journal]{IEEEtran}

\usepackage{amssymb, graphicx, cite, amsmath, psfrag, minibox, mathrsfs, bigints, stfloats, color, upgreek, gensymb, epstopdf, textcomp, amsthm, amsfonts, euscript, calligra, yhmath, cancel}

\usepackage[normalem]{ulem}

\usepackage[font=footnotesize]{subcaption}
\usepackage[font=footnotesize]{caption}
\usepackage{hyperref}

\usepackage{mathtools, amsfonts, multirow, array}
\usepackage{verbatim, mdwtab, relsize, cleveref, mdwmath}
\usepackage{algorithmicx}
\usepackage{algpseudocode}
\usepackage[ruled]{algorithm2e}
\usepackage{collectbox}
\usepackage{comment}

\usepackage{todonotes}

\usepackage{setspace}
\usepackage{float}

\usepackage{tikz}
\usetikzlibrary{shapes,arrows}
\usepackage{textcomp}

\makeatletter
\newcommand{\leqnomode}{\tagsleft@true}
\newcommand{\reqnomode}{\tagsleft@false}
\makeatother

\def\pioh{\widehat{\pi}_0}
\def\pilh{\widehat{\pi}_1}
\def\P0{P^{(0)}}

\def\Pbar{\overline{P}_{\rm{av}}}
\def\Ibar{\overline{I}_{\rm{av}}}
\def\SUtx{SU$_{\rm{tx}}$}
\def\SUrx{SU$_{\rm{rx}}$}
\def\PUrx{PU$_{\rm{rx}}$}

\def\Tf{T_{\rm{f}}}
\def\Tsen{T_{\rm{se}}}
\def\Nsen{N_{\rm{se}}}
\def\Ttra{T_{\rm{tr}}}
\def\Ptra{P_{\rm{tr}}}
\def\Ntra{N_{\rm{tr}}}

\def\mPU{m_{\rm{PU}}}
\def\mSR{m_{\rm{SR}}}

\def\phiPU{\phi_{\rm{PU}}}
\def\phiSR{\phi_{\rm{SR}}}

\def\hsp{h_{\rm{sp}}}
\def\gss{g_{\rm{ss}}}
\def\gsp{g_{\rm{sp}}}

\def\Pp{P_{\rm{p}}}
\def\Neq{N_{\rm{eq}}}
\def\RLB{R_{\rm{LB}}}
\def\Rora{R_{\rm{S}_1}}
\def\Rnew{R_{\rm{S}_2}}
\def\Nd{N_{\rm{d}}}
\def\Td{T_{\rm{d}}}
\def\Tsam{T_{\rm{s}}}
\def\Dd{D_{\rm{d}}}
\def\Dtr{D_{\rm{tr}}}

\def\Sigmaw{\sigma_{\rm{w}}}
\def\Sigmap{\sigma_{\rm{p}}}
\def\Sigmaq{\sigma_{\rm{q}}}

\ifCLASSINFOpdf
\else
\fi

\begin{document}

\title{Achievable Rates of Opportunistic Cognitive Radio Systems Using Reconfigurable Antennas with Imperfect Sensing and Channel Estimation}

\author{Hassan Yazdani,
	Azadeh Vosoughi,~\IEEEmembership{Senior Member,~IEEE}
	  Xun Gong,~\IEEEmembership{Senior Member,~IEEE}\\
        University of Central Florida\\
        E-mail: {\tt \normalsize h.yazdani@knights.ucf.edu, azadeh@ucf.edu, xun.gong@ucf.edu } 
}


\markboth{}%
{Shell \MakeLowercase{\textit{et al.}}: Bare Demo of IEEEtran.cls for Journals}
%

\newcounter{MYeqcounter}
\multlinegap 0.0pt                     

\maketitle

\vspace{-0mm}
\begin{abstract}	
We consider an opportunistic cognitive radio (CR) system in which secondary transmitter (\SUtx) is equipped with a reconfigurable antenna (RA). Utilizing the beam steering capability of the RA, we regard a design framework for integrated sector-based spectrum sensing and data communication. In this framework, \SUtx ~senses the spectrum and detects the beam corresponding to active primary user's (PU) location. \SUtx ~also sends training symbols (prior to data symbols), to enable channel estimation at secondary receiver (\SUrx) and selection of the strongest beam between \SUtx--\SUrx ~for data  transmission. We establish a lower bound on the achievable rates of \SUtx--\SUrx ~link, in the presence of spectrum sensing and channel estimation errors, and errors due to incorrect detection of the beam corresponding to PU's location and incorrect selection of the strongest beam for data transmission. We formulate a novel constrained optimization problem, aiming at maximizing the derived achievable rate lower bound subject to average transmit and interference power constraints. We optimize the durations of spatial spectrum sensing and channel training as well as data symbol transmission power. Our numerical results demonstrate that between optimizing spectrum sensing and channel training durations, the latter is more important for providing higher achievable rates.
\end{abstract}
%
%
\vspace{-0mm}

\IEEEpeerreviewmaketitle

\section{Introduction}\label{Se1}

\subsection{Literature Review}
Cognitive radio (CR) technology improves spectrum utilization and fills the spectral holes, via  allowing an unlicensed or secondary user (SU) to access licensed bands in a such way that its imposed interference on license holder primary users (PUs) is restricted \cite{Goldsmith2}. CR systems are mainly classified as  underlay CR and opportunistic (or interweave) CR systems. In underlay CR systems, SUs use a licensed frequency band simultaneously with PUs, as long as the interference caused by SUs and imposed on PUs stays below a pre-determined threshold \cite{Goldsmith2, ICASSPpaper, Joneidi}. While underlay CR systems do not require spectrum sensing to detect PUs' activities, they demand coordination between PUs and SUs to obtain channel state information (CSI), that is not always feasible. In opportunistic CR systems, SUs use a licensed frequency band during a time interval, only if  that frequency band is not used by PUs. While opportunistic CR systems do not require  coordination between PUs and SUs to acquire CSI corresponding to SU-PU link (and hence the system implementation is easier), they necessitate spectrum sensing to monitor and detect PUs' activities. 
\par The CR literature mainly assume that SU has access to full CSI of all links for its operation.  However, in practice, SU has access only to partial CSI,  due to several factors including channel estimation error, mobility of PU or SU, and limitation of  feedback channel. Partial (imperfect) CSI has deteriorating effects on the fundamental performance limits of CRs and should not be overlooked. We note that the impact of partial CSI on the performance of underlay and opportunistic CR systems are different, due to inherent distinctions between these two CR systems.  For underlay CR systems, several researchers have studied the impact of imperfect CSI on the ergodic capacity \cite{Alouini8, Xu2013, Shafi2, Alouini6, Shafi3, Sharma1} and symbol error probability \cite{Mehta2}. In particular, references \cite{Alouini8, Xu2013, Shafi2} focus on investigating  the impact of imperfect CSI of \SUtx--PU receiver (\PUrx) ~link on the optimal transmit power of \SUtx ~that maximizes the constrained capacity of \SUtx--\SUrx ~link, where \SUtx ~cannot always satisfy the interference power constraint (due to partial CSI) and has to reduce its transmit power. The authors in \cite{Alouini6, Shafi3} considered the impact of partial CSI for both \SUtx--\PUrx ~and \SUtx--\SUrx ~links on the CR system capacity. Different from \cite{Alouini6, Shafi3, Alouini8, Xu2013 , Shafi2}, \cite{Sharma1} discussed the trade-off between channel estimation accuracy and channel estimation duration (time). The authors in \cite{Sharma1} studied the optimal transmit power of \SUtx ~and optimal channel estimation duration, such that the capacity of \SUtx--\SUrx ~link is maximized, subject to a constraint on interference power imposed on \PUrx. 
\par In opportunistic CR systems, spectrum sensing is necessary for detecting PUs' activities and protecting the PUs against harmful interference. In general, any spectrum sensing (signal in noise detection) technique is prone to errors, that can be described as mis-detection or false alarm probability \cite{ AsilomarPaper, GlobalSIP}. On the other hand, imperfect CSI of \SUtx--\SUrx ~link due to channel estimation error (even under perfect spectrum sensing) has negative influence on the link capacity. Imperfect spectrum sensing exacerbates the negative effect of imperfect CSI on the link capacity. Hence, for opportunistic CR systems, one needs to study the combined impacts of imperfect spectrum sensing and imperfect CSI on the system performance. Such study presents new challenges, compared with studies that focus on understanding only the effect of imperfect spectrum sensing, when CSI is perfect (or vice versa), on the link capacity. To the best of our knowledge, there are only a few works that have considered the aforementioned combined effects in their system performance analysis \cite{Sharma2, Gursoy4, Gursoy7, Gursoy9}. For example in \cite{Sharma2} \SUtx ~estimates the received power from PU during sensing-estimation time and monitors PU's activity. If the spectrum is sensed idle, \SUtx ~with its imperfect CSI of \SUtx--\SUrx ~link, sends data to \SUrx ~with a fixed power. The authors showed that the constrained capacity of \SUtx--\SUrx ~link can be significantly enhanced (subject to a constraint on the detection probability), via optimizing sensing-estimation time. The authors in \cite{Gursoy4} considered a delay-sensitive CR system with a different setup, where after spectrum sensing at \SUtx, \SUtx ~transmits at fixed powers and rates, where these fixed values depend on the result of spectrum sensing (i.e., the transmit power and rate corresponding to spectrum being sensed idle are different from those corresponding to spectrum being sensed busy). The authors optimized these fixed powers and rates such that the defined effective capacity is maximized, subject to average transmit power and buffer length constraints. The authors in \cite{Gursoy7} considered a related problem to \cite{Gursoy4}, where the two data transmit power levels are given and instead two training power levels as well as  training period are optimized to maximize the achievable rate.  The work in \cite{Gursoy9} considered different levels of CSI corresponding to \SUtx--\SUrx ~and \SUtx--PU links, and studied optimal transmit power levels of \SUtx, such that the capacity of \SUtx--\SUrx ~link is maximized, where the optimized power levels depend on the level of CSI. 
\par In the above cited works SUs are equipped with single antenna. Multiple antennas and in particular transmit beamforming  techniques have been  utilized to ameliorate the performance degradation due to the interference imposed on PUs for underlay CR systems \cite{Poor1, Liang8, Mehta3} and opportunistic CR systems \cite{Gursoy10} with perfect CSI of \SUtx--\SUrx ~link available at \SUtx.  The authors in \cite{Ropokis} considered an opportunistic CR system, where \SUtx ~has a single antenna and \SUrx ~has multiple antennas and applies maximum ratio combining (MRC) technique, and studied the combined effects of spectrum sensing error and  imperfect CSI of \SUtx--\SUrx ~link at \SUtx ~on the system bit error rate (BER) performance. Optimal spectrum sensing time, channel estimation time, and \SUtx ~transmit power are obtained, such that BER is minimized, subject to average transmit and peak interference power constraints. We note that the benefits of multi-antenna techniques come at the cost of requiring  an expensive and power-hungry radio frequency (RF) chain per antenna, which consists of digital-to-analog converters, filters, mixers, and amplifiers. 
\par Alternatively, a reconfigurable antenna (RA), which has only one RF chain, is  a low-complexity and low-cost technology that addresses this challenge \cite{Gong1, Gong3, Gong4}.  RAs enable efficient exploitation of spatial diversity (via dynamically adjusting radiation pattern and beam steering/scanning capability) for reliable spectrum sensing and data transmission in CR systems. They are also capable of changing their parameters to dynamically adjust their polarization, carrier frequency and bandwidth \cite{Gong1, Gong3, MILCOM}. For both underlay and opportunistic  CR systems, RAs are used to increase  signal-to-noise ratio (SNR) for transmission and reception of directional signals \cite{Papadias2}, enhance spectrum sensing \cite{Papadias2, Max_min_SSS_Parasitic, Blind_Spatial_SS}, and limit interference to and from PUs \cite{J1, CISS2019}. Motivated by the advantages of RAs, in our study we assume that \SUtx ~is equipped with an RA that has beam steering capability. 
%
\subsection{Knowledge Gap, Research Questions, and Our Contributions}
To the best of our knowledge, our work is the first to consider the combined effects of spectrum sensing error and imperfect CSI of \SUtx--\SUrx ~link on the { achievable rates} of an opportunistic CR system with a RA at \SUtx. In our opportunistic CR system, \SUtx ~relies on the beam steering capability of RA to detect the direction of PU's activity and also to select the strongest beam for data transmission to \SUrx. We assume \SUtx ~sends training symbols to enable channel estimation at \SUrx, and employs Gaussian input signaling for transmitting its data symbols to \SUrx. Also, \SUrx ~shares its imperfect CSI of \SUtx--\SUrx ~link with \SUtx ~through an error-free low-rate feedback channel.
\par Assuming that there are average transmit power constraint (ATPC) and average interference constraint (AIC), we provide answers to the following research questions: How does spectrum sensing error affect accuracy of detecting the direction of PU's activity, estimating \SUtx--\SUrx ~channel, and selecting the strongest beam for data transmission? How do training symbol transmission and beam detection error (error in obtaining the true direction of PU's activity) affect interference imposed on PU? How do the combined effects of spectrum sensing error and channel estimation error, as well as beam detection error and beam selection error (error in finding the true  strongest beam for data communication to \SUrx) impact the achievable rates for reliable communication over \SUtx--\SUrx ~link?  How do the trade-offs between spatial spectrum sensing time, channel training time, data transmission time, training and data symbol transmission powers affect the achievable rates? How can we utilize these trade-offs to design transmit power control strategies, such that the achievable rates subject to ATPC and AIC are maximized? Our main contributions follow:

\noindent 1) Given this system model, we establish a lower bound on  the achievable rates of \SUtx--\SUrx ~link, in the presence of both spectrum sensing error and channel estimation error. We formulate a novel constrained optimization problem, aiming at maximizing the derived lower bound subject to AIC and ATPC.

\noindent 2) Our problem formulation takes into consideration the combined effects of imperfect spectrum sensing and channel estimation as well as the errors due to (i) incorrect detection  of the beam corresponding to PU's location (and its corresponding effect on average interference imposed on PU) occurred during spatial spectrum sensing phase, (ii) incorrect selection of the strongest beam for data transmission from \SUtx~ to \SUrx, occurred during channel training phase. These beam detection and beam selection errors are introduced by the RA at \SUtx.

\noindent  3) Given a fixed-length frame, we optimize the durations of spatial spectrum sensing and channel training as well as data symbol transmission power. Based on the structure of the optimized transmit power, we propose alternative power adaptation schemes that are simpler to implement and yield lower bounds on the achievable rates that are very close to the one produced by the optimized transmit power. 
%
%
\subsection{Paper Organization}
The remainder of the paper is organized as follows. Section \ref{SysModel} explains our system model consisting of three phases: spatial spectrum sensing phase, channel training phase, and data transmission phase. Sections \ref{Phase1} and \ref{Phase2} describe spatial spectrum sensing phase and channel training phase, respectively. Section \ref{Phase3} discusses data transmission phase, establishes a lower bound on the achievable rates, and characterizes ATPC and AIC. Then, it formalizes a constrained optimization problem with three optimization variables (durations of spatial spectrum sensing and channel training phases, and data symbol transmission power), aiming at maximizing the derived lower bound, subject to ATPC and AIC. Section \ref{Solving} provides solution to this constrained optimization problem. Section \ref{SimResults} presents our simulation results and Section \ref{Conclusion} concludes the paper. 
%
%
\section{System Model}\label{SysModel}
%
\begin{figure}[!t]
	\centering
	\includegraphics[width=45mm]{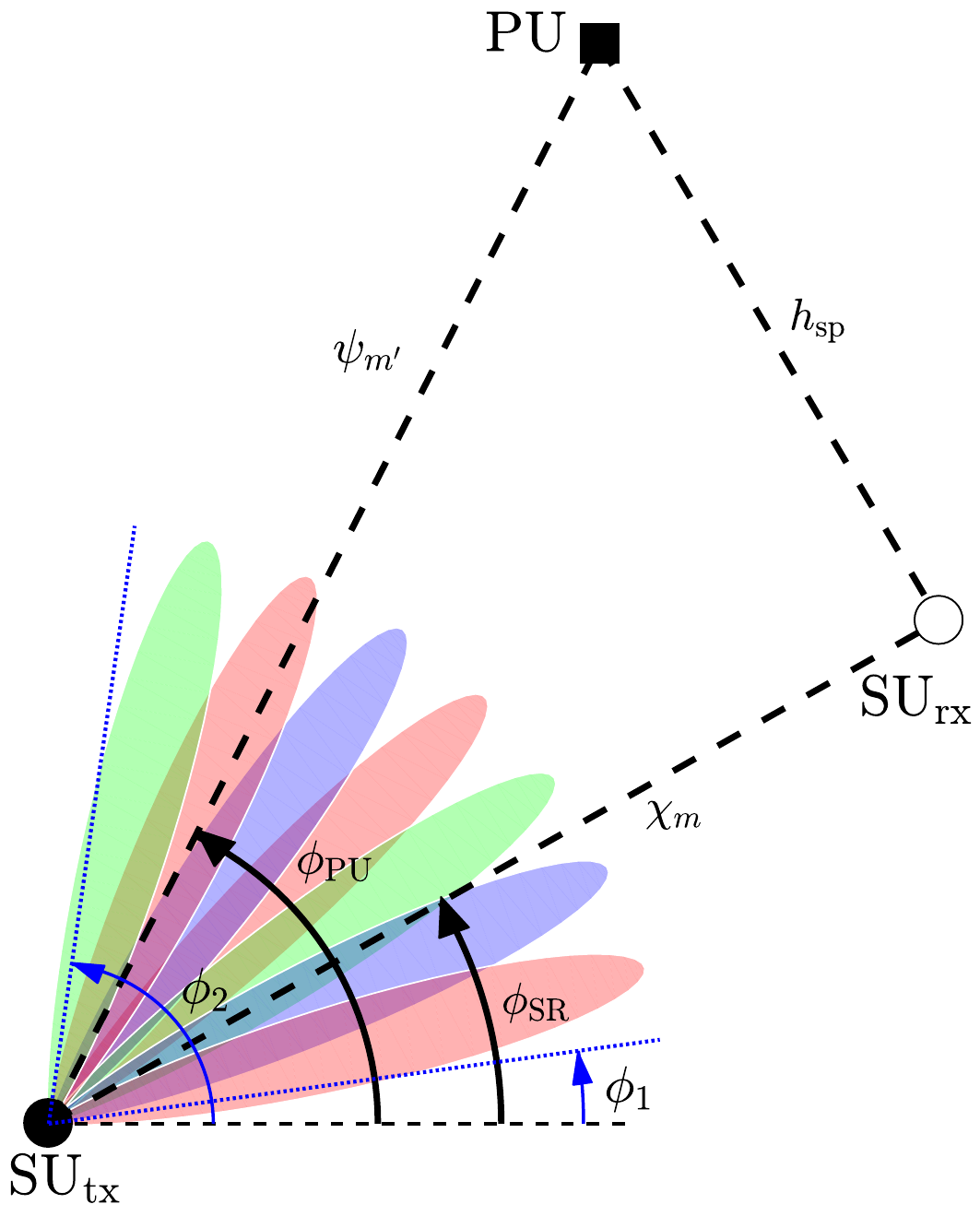}
	\caption{Our opportunistic CR system with an $M$-beam RA at \SUtx ~and omni-directional antennas at \SUrx ~and PU. } 
	\label{SystemModelFig} 
	\vspace{-0mm}  
\end{figure}
\subsection{Structure of a RA}
We consider a RA which can generate $M$	{\it beampatterns} and these beampatterns  cover the angular plane from $\phi_1$ to $\phi_2$, i.e., the angular space from $\phi_1$ to $\phi_2$ is divided into $M$ spatial {\it sectors} or beams\footnote{Throughout this paper, "sector" and "beam" are used interchangeably.}. 
One can extend this angular space to cover the entire azimuth plane.  The beampattern corresponding to $m$-th beam achieves its maximum at angle $\kappa_m= \frac{2 \pi(m-1)}{M}$ for $m=1, \ldots, M$. Fig. \ref{SystemModelFig} shows the beampatterns of a RA with $M=7$ beams. It is noteworthy that the RA can also reconfigure itself to generate an omni-directional pattern.
To mathematically model the radiation pattern of beams, we adopt the Gaussian pattern  in $x\!-\!y$ azimuth plane in terms of angle $\phi$ given by \cite{J1}
%
\begin{equation}
p(\phi) = \,A_1 + A_0 ~ e^{ -B \left ( \frac{ \mathcal{M(\phi)}} {\phi_{\rm{3dB}}} \right )^2 }, \qquad
\mathcal{M(\phi)} = \text{mod}_{2 \pi} ( \phi + \pi ) - \pi, 
\end{equation}
%
where $\text{mod}_{2 \pi} ( \phi  )$ denotes the remainder of $\frac{\phi}{2 \pi}$, $B=\ln(2)$, $\phi_{\rm{3dB}}$ is the 3-dB beamwidth, {$A_1$} and $A_0$ are two constant antenna parameters. The radiation pattern of $m$-th beam at angle $\phi$ is 
%
\begin{equation}\label{pm}
p_m(\phi ) = p(\phi - \kappa_m),  \qquad \text{for} ~m=1, \ldots, M. 
\end{equation}
%
In this paper, we discuss the received or transmitted signal at $m$-th beam of \SUtx. This implies that, during the signal reception or transmission, the \SUtx's antenna parameters are set and tuned  such that the beampattern corresponding to $m$-th beam is generated. Given the antenna design, we focus on how the sector-based structure of this RA can be exploited to enhance the system performance of our opportunistic CR system, in which \SUtx ~optimizes its sector-based data communication to \SUrx ~according to the results of its sector-based spectrum  sensing. 
%
\subsection{Description  of Our Opportunistic CR System}\label{Section2B}
Our opportunistic CR system model is illustrated in Fig. \ref{SystemModelFig}, consisting of a PU and a pair of \SUtx ~and \SUrx. We note that PU in our system model can be a primary transmitter or receiver. We assume when PU is active it is engaged in a bidirectional communication with another PU, which is located far from \SUtx ~and hence its activity does not impact our analysis. We assume \SUtx ~is equipped with an $M$-beam RA (for spatial spectrum sensing, channel training and data transmission) with the capability of choosing one out of $M$ sectors for its data transmission to \SUrx, while \SUrx ~and PU use omni-directional  antennas. We assume there is an error-free low-rate feedback channel\footnote{Given a low rate feedback, the error-free feedback channel is a reasonable assumption \cite{Liang8}.} from \SUrx ~to \SUtx, to enable \SUtx ~select the best sector for its data transmission to \SUrx, and to adapt its transmit power according to the \SUtx--\SUrx ~channel information. The direction (orientation) of PU and \SUrx ~with respect to \SUtx ~are denoted by angles $\phiPU$,  and $\phiSR$, receptively, where $\phiSR, \phiPU \in (\phi_1 , \phi_2)$. Clearly, in our problem \SUtx ~does not know these directions  or angles  (otherwise, the beam selection at \SUtx ~for data transmission would become trivial).
\par Let $h$, $h_{\rm{ss}}$, $\hsp$ denote the fading coefficients of channels between \SUtx ~and PU, \SUtx ~and \SUrx, and \SUrx ~and PU, respectively, when the RA of \SUtx ~is in omni-directional mode. We model these fading coefficients as independent circularly symmetric complex Gaussian random variables. Equivalently, $g \! =\! |h|^2$, $\gss \! = \!|h_{\rm{ss}}|^2$ and $\gsp \! =\! | \hsp|^2$ are independent  exponentially distributed random variables  with mean $\gamma$, $\gamma_{\rm{ss}}$ and $\gamma_{\rm{sp}}$, respectively\footnote{We note that the distances between users are included in the small scale fading model \cite{GoldSmith}, i.e., the mean values $\gamma, \gamma_{\rm{ss}}, \gamma_{\rm{sp}}$ encompass distance-dependent path loss.}. 
In our problem we assume that SUs and PU cannot cooperate, and hence SUs cannot estimate $g$ and $\gsp$. However, \SUtx ~knows the channel statistics, i.e., the mean values $\gamma$ and $\gamma_{\rm{sp}}$.  
Let $\psi_{m^\prime}$ and $\chi_m$ denote the fading coefficients of channel between $m^\prime$-th sector of \SUtx ~and PU, and between $m$-th sector of \SUtx ~and \SUrx, respectively, when the RA of \SUtx ~is in directional mode. Using the radiation pattern expression in \eqref{pm} we can relate $\psi_{m^\prime}$ to $h$ and $\chi_m$ to $h_{ \rm{ss}}$ as  $\psi_{m^\prime} \! =\! h  \sqrt{ p_{m^\prime} (\phiPU )} $, $\chi_m \!= \! h_{ \rm{ss}} \sqrt{p_m( \phiSR )}$. 
We assume the channel gain $\nu_m = |\chi_m|^2$ is an exponentially distributed random variable with mean $\alpha_m$, and \SUtx ~knows $\alpha_m$, for all $m$ \cite{ArslanESPAR, J1}. For the readers' convenience, we have collected the most commonly used symbols in Table \ref{table1}.
%
%
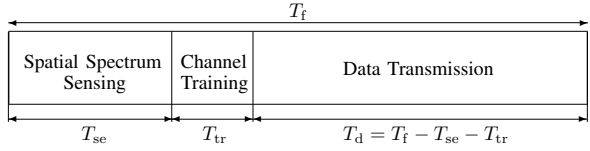
\begin{figure}[!t]
\vspace{-8mm}
\centering
\hspace{0mm}
\setlength{\unitlength}{3.2mm} 
\centering
\scalebox{0.75}{
\begin{picture}(27,9)
\hspace{0mm}
\centering
\put(-2.55,-0.5){\line(0,1){5}}
\put(-2.5,.5){\framebox(32,4){}}
\put(-1.7,2.5){Spatial Spectrum}
\put(0.5,1.3){Sensing}
\put(6.5,-0.5){\line(0,1){5}}
\put(7.0,2.5){Channel}
\put(7.0,1.3){Training}
\put(11.0,-0.5){\line(0,1){5}}
\put(16,2.1){Data Transmission}
\put(29.52,-0.5){\line(0,1){5}}
\put(-2.55,5.0){\vector(1,0){32.1}}
\put(-2.55,5){\vector(-1,0){0}}
\put(13,5.3){$\Tf$}
\put(-2.55,-0.35){\vector(1,0){9.1}}
\put(-2.55,-0.35){\vector(-1,0){0}}
\put(1.5,-1.5){$\Tsen$}
\put(6.5,-0.35){\vector(1,0){4.5}}
\put(6.5,-0.35){\vector(-1,0){0}}
\put(8.2,-1.5){$\Ttra$}
\put(11.0,-0.35){\vector(1,0){18.5}}
\put(11.0,-0.35){\vector(-1,0){0}}
\put(16,-1.5){$  \Td = \Tf - \Tsen - \Ttra$}
\end{picture}}
\vspace{4.5mm} 
\caption{The structure of frame employed by \SUtx.} 
\label{FrameStructure}
\vspace{-2mm} 
\end{figure}
%
%
\begin{table}[h!]
	\begin{center}
		\caption{Most commonly used symbols.}
		\label{table1}
		\begin{footnotesize}
			\begin{tabular}{l|l} 
			\textbf{Symbol} & \textbf{Description} \\
			\hline \hline
			$M$ & Number of beams\\
			$\Nsen$ & Number of samples used for {\it spatial spectrum sensing} \\
			$\Ntra$ & Number of samples used for {\it channel training} \\
			$\Ptra$ & Power of training symbols \\ 
			$\psi_{m^\prime}$ &  Fading coefficient of channel between $m^\prime$-th beam of \\ & \SUtx ~and PU \\                        
			$\chi_m, \widehat{\chi}_m, \widetilde{\chi}_m$ &  Fading coefficient of channel between $m$-th beam of \\ & \SUtx ~and \SUrx, LMMSE channel estimate, and its \\ & corresponding estimation error \\  
			$\alpha_m, \widehat{\alpha}_m, \widetilde{\alpha}_m$ &  Variances of $\chi_m, \widehat{\chi}_m, \widetilde{\chi}_m$ \\  
			$\mPU^*, \mSR^*$ \!\!\!\!\! & Indices of selected beam for  PU and \SUrx \\
			$\widehat{\nu}^*$ &  Channel gain of selected beam for data transmission \\ & from  \SUtx ~to \SUrx \\       
			\end{tabular}
		\end{footnotesize}   
	\end{center}   
\vspace{-5mm}  
\end{table}
\par Suppose, SUs employ a frame with a fixed duration of $\Tf$ seconds, depicted in Fig. \ref{FrameStructure}. We assume the \SUtx--\SUrx ~channel remains constant over the frame duration. \SUtx ~first senses the spectrum and monitors PU's activity. We refer to this period as {\it{spatial spectrum sensing }} phase with a variable duration of $\Tsen = M \Nsen \Tsam$ seconds, where $\Tsam$ is the sampling period and $\Nsen$ is the number of collected samples during this phase per beam. Suppose $\mathcal{H}_1$  and $\mathcal{H}_0$  represent the binary hypotheses of PU being active and inactive, respectively, with prior probabilities $\Pr \{\mathcal{H}_1\} \! = \!  \pi_1$ and $\Pr \{\mathcal{H}_0\} \! =\! \pi_0$. \SUtx ~applies a binary detection rule to decide whether or not PU is active. The details of the binary detector are presented in Section \ref{Eigenvalue_Detector}. While being in this phase, \SUtx ~determines the beam corresponding to the orientation of  PU based on the received signal energy as we describe in Section \ref{PU_Beam_Selection}. 
\par Depending on the outcome of spectrum sensing, \SUtx ~stays in {spatial spectrum sensing phase}  or enters the next phase, which we refer to as {\it{channel training phase}} with a variable duration of $\Ttra = M \Ntra \Tsam$  seconds. 
In this phase, \SUtx ~sends $\Ntra$ training symbols with fixed symbol power $\Ptra$ per beam to enable channel estimation at \SUrx, as we explain in Section \ref{Ch_Training}.
Based on the results of channel estimation for all beams, \SUrx ~selects the beam with the largest \SUtx--\SUrx ~fading gain, as we describe in Section \ref{SubSe3}. 
This information as well as the  corresponding beam index are shared with \SUtx ~via the feedback channel. 
Next, \SUtx ~enters {\it data transmission phase} with a variable duration of $ \Td = \Tf \! - \! \Tsen \! - \! \Ttra$ seconds. During this phase, \SUtx ~sends $ \Nd = \Td / \Tsam$ Gaussian data symbols with adaptive symbol power $P$ to \SUrx ~over the selected strongest beam.  \SUtx ~adapts $P$ aiming at maximizing the achievable rates, subject to ATPC and AIC as we describe  in Section \ref{Phase3}. In the following sections, we describe how \SUtx ~operates during {spatial spectrum sensing phase}, {channel training phase}, and {data transmission phase}.
%
\section{Spatial Spectrum Sensing Phase}\label{Phase1}

\subsection{ Eigenvalue-Based Detector for Spatial Spectrum Sensing } \label{Eigenvalue_Detector} 
Let $\widehat{\mathcal{H}}_1$ and $\widehat{\mathcal{H}}_0$ denote the detector outcome, i.e., the detector finds PU active (spectrum is sensed busy and occupied) and inactive (spectrum is sensed idle and unoccupied and thus can be used by \SUtx ~for data transmission), respectively. Suppose when PU is active, it transmits signal $s(t)$ with power $\Pp$. Let $y_m(n)$ denote the discrete-time representation of received signal at $m$-th sector of \SUtx ~at time instant $t=n \Tsam$. 
We model PU's transmitted signal $s(n)$ as a zero-mean complex Gaussian random variable with variance $\Pp$ and we assume \SUtx ~knows $\Pp$. Since \SUtx ~collects $\Nsen$ samples per beam during spatial spectrum sensing phase, the hypothesis testing problem at discrete time instant $n$ for $m$-th sector is
%
\begin{equation}
\begin{aligned} 
{\cal H}_{0}\!: & ~~~y_m(n) =  w_m(n), \\
{\cal H}_{1}\!: & ~~~y_m(n) =  \psi_m(n)  s(n) + w_m(n). 
\end{aligned}
\end{equation}
%
%
%
%
The term $w_m(n)$ is the additive noise at $m$-th sector of \SUtx ~antenna and is modeled as 
$w_m(n) \sim \mathcal{CN}(0,\Sigmaw^2) $. We assume that $\psi_m(n)$, $s(n)$ and $w_m(n)$ are mutually independent random variables. Since \SUtx ~takes samples of the received signal for different sectors sequentially (in different time instants), $\psi_m(n)$ and $w_m(n)$ are independent and thus uncorrelated  both in time and space (sector) domains. Under hypothesis $\mathcal{H}_1$, given $\psi_m$, we have $y_m(n) \sim \mathcal{CN}( 0, \sigma_m^2 \!+\! \Sigmaw^2)$ where $\sigma^2_{m} = |\psi_m|^2 \Pp$. Under  hypothesis $\mathcal{H}_0$, we  have $y_m(n)  \sim \mathcal{CN} (0,\Sigmaw^2)$. 
%
%
\par Our proposed binary detector uses all the collected samples from $M$ sectors. To facilitate the signal processing needed for the binary detection, we define an $M \times \Nsen$ sample matrix $\boldsymbol{Z}=[\boldsymbol{z}_1, \ldots , \boldsymbol{z}_{\Nsen}]$, where the first row of $\boldsymbol{Z}$ is the $\Nsen$ samples collected from the first sector, the second row of $\boldsymbol{Z}$ is the $\Nsen$ samples collected from the second sector, and so forth. Given our assumptions, the columns of $\boldsymbol{Z}$ are orthogonal under both hypotheses, that is
%
\begin{align}
\mathbb{E} \big \{\boldsymbol{z}_i \boldsymbol{z}_j^H  & | \mathcal{H}_0 \big \} =\boldsymbol{0 },  ~~~~~~~ \mathbb{E} \big \{\boldsymbol{z}_i \boldsymbol{z}_j^H |\mathcal{H}_1 \big \} =\boldsymbol{0}, \nonumber \\
& \text{for} ~i \neq j, ~~~i, j=1, \ldots ,\Nsen
\end{align}
%
%
\noindent where  $ \mathbb{E} \{ \cdot \}$ is the statistical expectation operator and have the below covariance matrices 
%
\begin{subequations}
	\begin{align}
	\boldsymbol{\Gamma}_0 = & \mathbb{E} \left \{  \boldsymbol{z}_j  \boldsymbol{z}^H_j | \mathcal{H}_0  \right \} = \Sigmaw^2 \boldsymbol{ I}_M,  \\
	\boldsymbol{\Gamma}_1 = & \mathbb{E} \left \{  \boldsymbol{z}_j  \boldsymbol{z}^H_j | \mathcal{H}_1 ,  \boldsymbol{\psi} \right \} = \Pp \boldsymbol{\psi} \boldsymbol{\psi}^H  + \Sigmaw^2 \boldsymbol{ I}_M, 
	\end{align}
\end{subequations}
%
%
\noindent where vector $\boldsymbol{\psi} = \big [ \psi_1, \psi_2,  \ldots, \psi_M \big ]^T$. Therefore the sample covariance matrix $\widehat{\boldsymbol{ R}}$ becomes $\widehat{\boldsymbol{ R} } = \frac{1}{\Nsen } \boldsymbol{Z} \boldsymbol{Z}^H. $
%
%
%

\noindent Let $f({\bf Z} | {\cal H}_{0})$ and $f({\bf Z} | {\cal H}_{1}, \boldsymbol{\psi} )$ denote the probability distribution function (pdf) of  ${\bf Z}$ under ${\cal H}_{0}$ and ${\cal H}_1$ (given $\boldsymbol{\psi}$), respectively. These pdf expressions are
%
\begin{subequations}
	\begin{align}
	f({\bf Z} | {\cal H}_{0}) \! =  \, & {1\over (\pi \Sigmaw^2)^{ \Neq }}\exp \left\{ \! {{ {\rm tr}({\bf ZZ}^{H})} \over{ -\Sigmaw^2}} \!\right\},  \\
	f({\bf Z}|{\cal H}_{1}, \boldsymbol{\psi}) \! =  \, & {1\over {\pi^{\Neq }\det({\bf \Gamma_1})^{ \Nsen }}} \exp \!\left\{ \! {{ {\rm tr}({\bf \Gamma}_1^{-1}{\bf ZZ}^{H})} \over{ -\Sigmaw^2 }} \!\right\},
	\end{align}
\end{subequations}
%
where $\Neq = M \Nsen$. The optimal detector would compare the logarithm of likelihood ratio (LLR) against a threshold $\eta_0$ to detect the PU's activity as below
%
%
\begin{equation}\label{LLRT}
\text{LLR} = \ln \frac{f({\bf Z}|{\cal H}_{1}, \boldsymbol{\psi}) }{f({\bf Z} | {\cal H}_{0})} \gtreqless
\begin{matrix}
\widehat{\cal H}_{1} \cr \widehat{\cal H}_{0}
\end{matrix}
\eta_0.
\end{equation}
%
%
%
In the absence of the knowledge of the fading coefficients vector $\boldsymbol {\psi}$, \SUtx ~obtains the generalized likelihood ratio test (GLRT) \cite{Awin_Tutorial, Liang3, Liang5, Gazor, Ratnarajah} which uses the maximum likelihood (ML) estimate of $\boldsymbol{\psi}$ under $\mathcal{H}_1$. Let $\mathcal{L}_1 ({\bf Z}) = \ln f({\bf Z}|{\cal H}_{1}, \boldsymbol{\psi})$. To find the maximum of $\mathcal{L}_1 ({\bf Z})$ with respect to $\boldsymbol{\psi}$, we take the derivative of $\mathcal{L}_1 ({\bf Z})$ with respect to $\boldsymbol{\psi}$ and  solve $\frac{\partial}{\partial \boldsymbol{\psi}} \mathcal{L}_1 ({\bf Z}) = {\bf 0}$ for $\boldsymbol{\psi}$. The obtained solution is the ML estimate of $\boldsymbol{\psi}$. Substituting  this solution into \eqref{LLRT} and after some mathematical manipulation, we reach the following decision rule $T =\frac{ \lambda_\text{max} }{\Sigmaw^2 }  \gtreqless
{\small
	\begin{matrix}
	\widehat{\cal H}_1 \cr \widehat{\cal H}_0
	\end{matrix} }
\eta$ \cite{Gazor},
%
%
\noindent where $T$ is the test statistics, $\lambda_\text{max} $ is the maximum eigenvalue of $\widehat{\boldsymbol{ R}}$, and $\eta$ is the threshold.  
For large $\Nsen$, $ T$  under ${\cal H}_0$ is distributed as Tracy-Widom distribution of order 2 \cite[Lemma~1]{Gazor} and the probability of false alarm $P_{\rm{fa}} = \Pr ( \widehat{\cal H}_1 | {\cal H}_0) = \Pr( T > \eta | {\cal H}_0 )$ is
%
\begin{equation}\label{Prfa}
P_{\rm{fa}} = 1-F_\text{TW2} \left( \frac{\eta \! -\! \theta_{\rm{sen}} }{\sigma_{\rm{sen}} } \right ),
\end{equation}
%
\noindent where $F_\text{TW2} (\cdot)$ is the commutative distribution function (CDF) of Tracy-Widom distribution of order 2 and $\theta_{\rm{sen}}$ and $\sigma_{\rm{sen}}$ in \eqref{Prfa} are given below
%
\begin{subequations}
	\begin{align}
	&\theta_{\rm{sen}} =  \bigg (1 + \sqrt{\frac{M}{\Nsen}} \bigg )^2, \\
	\sigma_{\rm{sen} } =  \frac{1}{\sqrt{\Nsen}} &  \bigg (1 \!+ \! \sqrt{\frac{M}{\Nsen}} \bigg )  \bigg (\frac{1}{\sqrt{\Nsen}} \!+\! \frac{1}{\sqrt{M}} \bigg ) ^{\frac{1}{3}}.
	\end{align}
\end{subequations}
%
%
%
For large $\Nsen$,  $T$ under ${\cal H}_1$  is Gaussian distributed \cite[Lemma~2]{Gazor} and the probability of detection $P_{\rm{d}} = \Pr ( \widehat{\cal H}_1 | {\cal H}_1 ) = \Pr( T > \eta | {\cal H}_1 )$ is \cite{Ratnarajah, Gazor}
%
\begin{equation}\label{Pd_g}
P_{\rm{d}} = Q \left(  \frac{\eta \sqrt{\Nsen}}{1 \!+ \! \delta_{\rm{sen}} } \! - \! \frac{M\!-\!1}{  \delta_{\rm{sen}} \sqrt{\Nsen }} \! - \! \sqrt{\Nsen } \right ),
\end{equation}
%
\noindent where  $  \delta_{\rm{sen}} = \frac{\Pp \| \boldsymbol{\psi} \|^2}{\Sigmaw^2}$. The average detection probability $\overline{P}_{\rm{d}}$ can be computed by averaging \eqref{Pd_g}  over vector $\boldsymbol {\psi}$, $\overline{P}_{ \rm{d}}= \mathbb{E}_{ \boldsymbol {\psi} } \{ P_{\rm{d}} \}$.
For a given $\overline{P}_{\rm{d}}$, we can numerically find $\eta$ and obtain $\overline{P}_{\rm{fa}}$ using \eqref{Prfa}.
%
%
%
We can also compute the probabilities of events $\widehat{ \mathcal{H}}_0$ and $\widehat {\mathcal{H}}_1$ as $\pioh \! =\!  \Pr \{ \widehat{ \mathcal{H}}_0 \}  =  \beta_0 + \beta_1 $ and $\pilh \! = \! \Pr \{ \widehat{ \mathcal{H}}_1 \}  =  1 - \pioh $, respectively, where
%
\begin{subequations} \label{Beta}
	\begin{align}
	\beta_0 = \Pr \{ \mathcal{H}_0 , \widehat{ \mathcal{H}}_0\}  = \pi_0 (1  -  \overline{P}_{\rm{fa}}), \\
	\beta_1  =  \Pr \{\mathcal{H}_1 ,\widehat{ \mathcal{H}}_0\}  =  \pi_1 (1  - \overline{P}_{\rm{d}}).
	\end{align}
\end{subequations}
%
\subsection{ Determining the Beam Corresponding to PU Direction}\label{PU_Beam_Selection}
During spatial spectrum sensing phase when the spectrum  is sensed busy, \SUtx ~determines the beam corresponding to the direction of PU based on the received signal energy. Let $\varepsilon_m$ be the energy of received signal at $m$-th beam. We have
%
\begin{equation}
\varepsilon_m = \frac{1}{\Nsen}  \sum_{n=1 + (m\!-\!1)\Nsen}^{m\Nsen} \!\!\!  \big | y_m(n) \big |^2. 
\end{equation}
%
\SUtx ~determines the beam with the largest amount of received energy $\mPU^* = \arg\max \{ \varepsilon_m \}$ among all beams. For large $\Nsen$, we invoke central limit theorem (CLT) to approximate $\varepsilon_m$'s as Gaussian random variables under both hypotheses. 
Thus, under ${\cal H}_0$ we approximate $\varepsilon_m$ as a Gaussian with distribution $\varepsilon_m \sim  \mathcal{N}  (\Sigmaw^2 , \Sigmaw^4/\Nsen)$. Similarly, under ${\cal H}_1$, given $\phiPU$  we approximate $\varepsilon_m$ as another Gaussian with distribution $\varepsilon_m \sim  \mathcal{N}  (\varrho_m, \sigma^2_{\varepsilon_m | {\cal H}_1})$, where the mean $\varrho_m = \gamma \Pp  \,p_m(\phiPU ) + \Sigmaw^2$, and the variance  $\sigma^2_{\varepsilon_m | {\cal H}_1}$ is given below
\vspace{-1mm}
\begin{equation}\label{sigma2_H1}
\sigma^2_{\varepsilon_m |{\cal H}_1 } =  \frac{1}{\Nsen } \Big [ \Sigmaw^4  + 3 \Pp^2  \gamma^2 \,p_m^2(\phiPU)  + 2 \Sigmaw^2 \Pp  \gamma  \,p_m(\phiPU) \Big ].
\end{equation}
%
%
\noindent We note that, there is a non-zero error probability when \SUtx ~determines the beam index $\mPU^*$, i.e., it is possible that $\mPU^*$ is not the true beam index corresponding to PU direction. 
\par
\begin{figure}[!t] 
\centering
	\begin{subfigure}[b]{0.25\textwidth}                
		\centering	         
		\includegraphics[width=43mm]{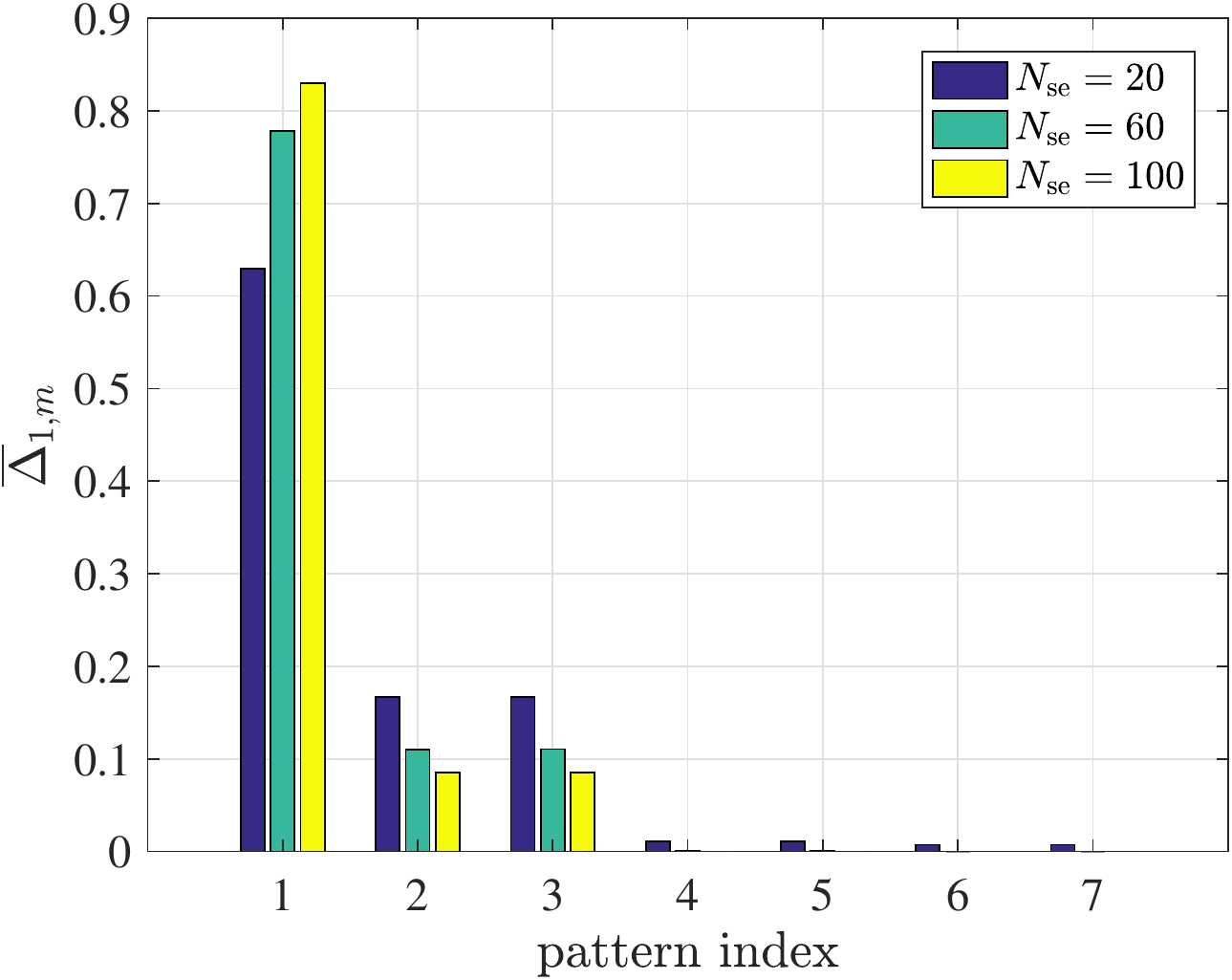}
		\caption{} 
		\label{PUSelection_Bar1}                   
	\end{subfigure}%
      \begin{subfigure}[b]{0.25\textwidth}
 		\centering      
		\includegraphics[width=43mm]{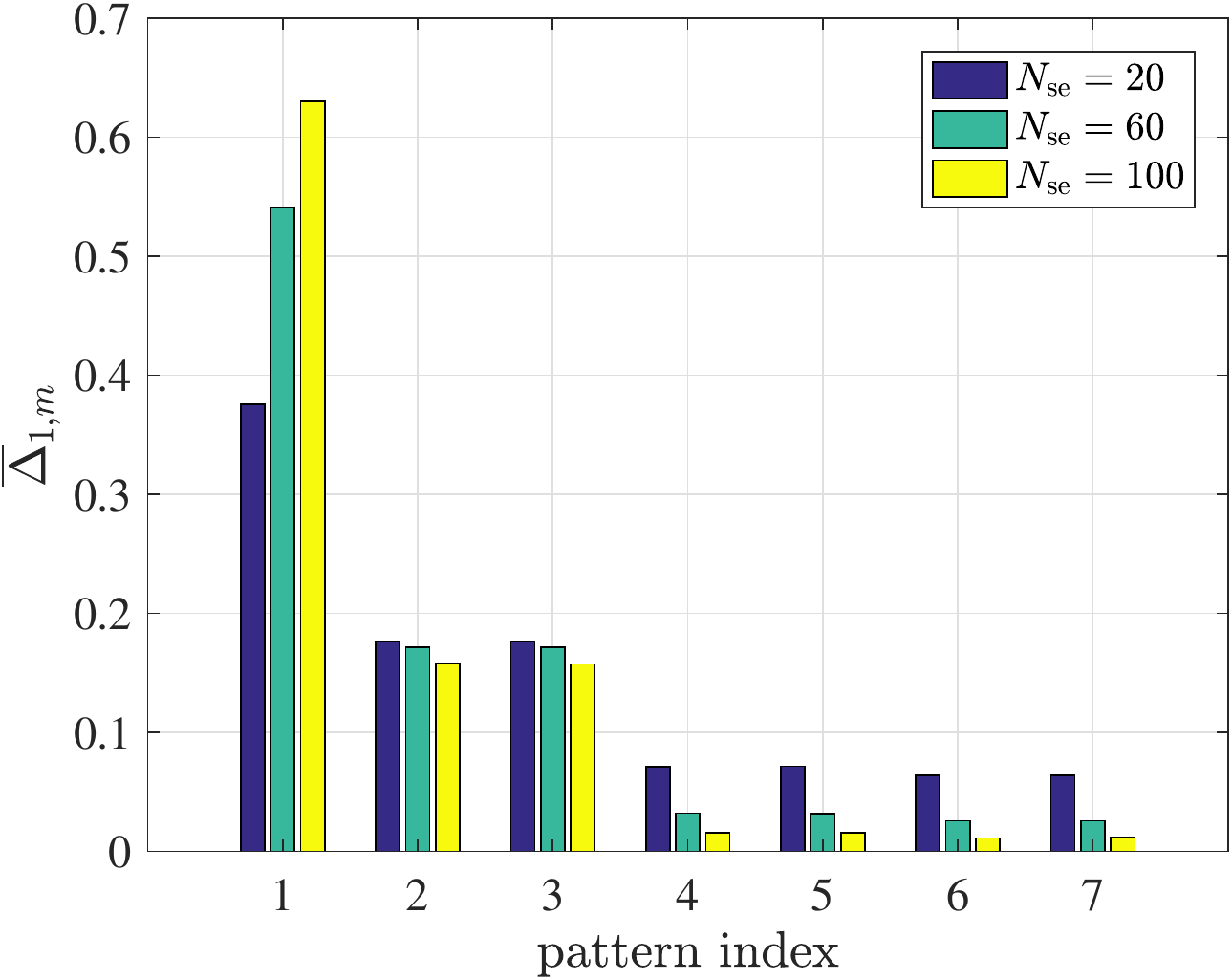}
		\caption{} 
		\label{PUSelection_Bar2}    
      \end{subfigure} \\
\caption{$\overline{\Delta}_{1,m}$ versus the index beam $m$ for $\phi_{\rm{3dB}}\!=\!20^\degree$ (a) $\text{SNR}_{\rm{PU}} \!= \!0$\,dB, (b) $\text{SNR}_{\rm{PU}} \!= \! -5$\,dB.}
\label{FigPUSelectionAll}
\end{figure}

 Let $\overline{\Delta}_{i,m}$ represent the average error probability of finding the sector index corresponding to PU direction, i.e., the probability that $\mPU^* =i$ while the true  PU direction lies in the angular domain of $m$-th sector, $\phiPU \in \Phi_m = \big [\frac{2\pi(m-3/2)}{M} ,\frac{2\pi(m-1/2)}{M}  \big )$, for $i \neq m, i, m =1, \ldots ,M$.  To find $\overline{\Delta}_{i,m}$ we start with finding $\Delta_i= \Pr\{ \mPU^* \!=\! i | \phiPU, \widehat{\cal H}_1 \}$, which is the probability that the index of selected sector is $i$, given $\phiPU$ and $\widehat{\cal H}_1$ (the binary detector in Section \ref{Eigenvalue_Detector} finds PU active). Note that under both hypotheses, $\varepsilon_m$'s are independent. Also, under ${\cal H}_0$, $\varepsilon_m$'s are identically distributed. Therefore, we have
%
\begin{align}\label{ProbSelection}
\Delta_i   
=  & \,\Pr \Big \{  \underset{ \forall m, \,m \neq i } {\varepsilon_i > \varepsilon_{m} } \,\big | \phiPU, \widehat{\cal H}_1 \Big \} \nonumber   \\ 
=  & \,\varsigma_1 \int_{0}^{\infty}  f_{\varepsilon_i | {\cal H}_1} \big (y  | \phiPU \big) \! \prod_{\underset{m \neq i} {m=1} }^{M} \! F_{\varepsilon_m |  {\cal H}_1 } \big (y  | \phiPU \big) dy \nonumber \\
+ & \,\varsigma_0 \int_{0}^{\infty}  f_{\varepsilon_m | {\cal H}_0} \big (y \big) F^{M-1}_{\varepsilon_m |{\cal H}_0 } \big (y \big) dy
\end{align}
%
where $f_{\varepsilon_m |{\cal H}_\ell }(x)$ and $F_{\varepsilon_m |{\cal H}_\ell }(x)$ are the pdf and CDF expressions of $\varepsilon_m$ under  ${\cal H}_\ell, \ell = 0,1$ and 
%
%
\begin{subequations}
	\begin{align}
	\varsigma_0 = \Pr \{ \mathcal{H}_0 | \widehat{ \mathcal{H}}_1\}  = {\pi_0 \overline{P}_{\rm{fa}} \over \pilh}, \\
	\varsigma_1  =  \Pr \{\mathcal{H}_1 |\widehat{ \mathcal{H}}_1\}  = {\pi_1  \overline{P}_{\rm{d}} \over \pilh}.
	\end{align}
\end{subequations}
%
Using $\Delta_i$, we find $\overline{\Delta}_{i,m}$ as the following
%
\begin{equation}\label{DeltaAverage}
\overline{\Delta}_{i,m} =  \int \nolimits_{ \phiPU \in \Phi_m }  \!\!\!\!\! \Delta_i \Pr \! \big \{ \phiPU \! \in \! \Phi_m \big \} ~d\phiPU. 
\end{equation}
%
%
Note that $\overline{\Delta}_{i,i}$ is the probability of selecting the correct beam and $\overline{\Delta}_{i,m}$ for $i \neq m$ is the probability of selecting the incorrect beam, leading to error probability in beam selection. The average error probability $\overline{ \Delta}_{1,m}$ versus the index beam $m$ is shown in Figs. \ref{PUSelection_Bar1} and \ref{PUSelection_Bar2} for $\text{SNR}_{\rm{PU}}\! = \! \gamma \Pp/\Sigmaw^2 \!=\!0, -5$\,dB. As expected, $\overline {\Delta}_{1,1}$ increases and $\overline {\Delta}_{1,m} , m \neq1$ decreases as $\Nsen$ increases.
%
%
%
%
%
\section{Channel Training Phase}\label{Phase2}
%
\subsection{ Channel Estimation at \SUrx}\label{Ch_Training}
During this phase, \SUtx ~sends the training vector $\boldsymbol{x}_t$ over all beams to enable channel estimation at \SUrx. Without loss of generality, we assume $\boldsymbol{x}_t= \sqrt{\Ptra} \,\boldsymbol{1}$, where $\boldsymbol{1}$ is an $\Ntra \times 1$  all-ones vector and $\Ptra$ is given. Let $ \boldsymbol{r}_m = \big [r_m(1) , \ldots, r_m(\Ntra) \big ]^T$ denote the discrete-time representation of received training symbols at \SUrx ~from $m$-th sector of \SUtx. We note that \SUtx ~enters this phase when the outcome of the binary detector in Section \ref{Eigenvalue_Detector} is $\widehat{\cal H}_0$. Due to error in spatial spectrum sensing, we need to differentiate the signal model for $r_m$ under ${\cal H}_0$ and ${\cal H}_1$. Assuming the fading coefficient $\chi_m$ is unchanged during the frame, we have 
%
%
%
%
\begin{equation} 
\begin{aligned}
{\cal H}_0, \widehat{\cal H}_0\!: & ~~~r_m(n) =  \chi_m \,\sqrt{\Ptra} + { q_m(n)}, \\
{\cal H}_1, \widehat{\cal H}_0\!: & ~~~r_m(n) = \chi_m \,\sqrt{\Ptra} + { \hsp (n)}  \,s(n) + { q_m(n)},
\end{aligned}
\end{equation}
%
where $q_m(n)$ is the additive noise at \SUrx ~antenna and is modeled as $q_m(n) \sim {\cal CN} \big (0,\Sigmaq^2 \big )$. The linear minimum mean square error (LMMSE) estimation  of fading coefficient ${\chi}_m$ when the spectrum sensing result is $\widehat{\cal H}_0$ can be obtained as \cite{kay93}
\vspace{-1mm}
\begin{subequations}
	\begin{align}
	\widehat{\chi}_m =\,& C_{\chi_m \boldsymbol{r}_m} C^{-1}_{\boldsymbol{r}_m} \,\boldsymbol{r}_m, \label{Xhat} \\
	C_{\chi_m \boldsymbol{r}_m} =\,& \mathbb{E} \{\chi_m \boldsymbol{r}^H_m | \widehat{\cal H}_0 \}  = \sqrt{\Ptra} \,\alpha_m \,\boldsymbol{1},\\
	C_{\boldsymbol{r}_m} =\,& \mathbb{E} \big \{\boldsymbol{r}_m \boldsymbol{r}^H_m | \widehat{\cal H}_0 \big \} = \omega_0 \,\mathbb{E} \big \{\boldsymbol{r}_m \boldsymbol{r}^H_m | {\cal H}_0, \widehat{\cal H}_0 \big \} \nonumber \\ 
	& \qquad \qquad \qquad~ +  \omega_1 \,\mathbb{E} \big \{\boldsymbol{r}_m \boldsymbol{r}^H_m | {\cal H}_1, \widehat{\cal H}_0 \big \} ,
	\end{align}	
\end{subequations}
%
\vspace{-0mm}
where
\vspace{-0mm}
%
\begin{subequations}
	\begin{align}
	\omega_0 = \Pr \{ \mathcal{H}_0 | \widehat{ \mathcal{H}}_0\}  = {\pi_0 (1 -  \overline{P}_{\rm{fa}} ) \over \pioh} = {\beta_0 \over \pioh}, \\
	\omega_1  =  \Pr \{\mathcal{H}_1 |\widehat{ \mathcal{H}}_0\}  = {\pi_1 ( 1 - \overline{P}_{\rm{d}}) \over \pioh} = {\beta_1 \over \pioh}.
	\end{align}
\end{subequations}
%
Finally, the LMMSE estimation of ${\chi}_m$ when the spectrum sensing result is $\widehat{\cal H}_0$, given in \eqref{Xhat}, reduces to 
\vspace{-0mm}
\begin{equation}\label{chi_hat}
\widehat{\chi}_m = \frac{\alpha_m \sqrt{\Ptra}  }{ \alpha_m  \Ptra \Ntra + { \Sigmaw^2 } + \omega_1 { \Sigmap^2 } } \sum_{n=1}^{\Ntra } r_m(n),
\end{equation}
%
%
\noindent where {$\Sigmap^2 = \Pp \gamma_{\rm{sp}}$}. The estimation error is $\widetilde{\chi}_m = \chi_m - \widehat{\chi}_m$ where $\widehat{\chi}_m$ and $\widetilde{ \chi}_m$ are orthogonal random variables \cite{kay93}, and $\widehat{\chi}_m$ and $\widetilde{\chi}_m$ are zero mean. Approximating  $\hsp(n) s(n)$ as a zero-mean Gaussian random variable with variance ${ \Sigmap^2 }$, we find that the estimate $\widehat{\chi}_m$ is distributed as a Gaussian mixture random variable \cite{Gursoy9, Gursoy5}. Let $\widehat{ \alpha}_m$ and $\widetilde{ \alpha}_m$, represent the variances of $\widehat{\chi}_m$ and $\widetilde{\chi}_m$, respectively. Also, Let $\widehat{ \alpha}^0_m$ and $\widehat{ \alpha}^1_m$ represent the variances of $\widehat{\chi}_m$ under ${\cal H}_0$ and ${\cal H}_1$, respectively. We have
\vspace{-1mm}
\begin{subequations}\label{alpha_hat}
	\begin{align}
	\widehat{ \alpha}^0_m \! =\! & \,\mathbb{VAR} \{\widehat{\chi}_m | {\cal H}_0 ,{ \widehat{\cal H}_0}  \} \! =\!  \frac{\alpha_m^2 \Ptra \Ntra \big (\alpha_m \Ptra \Ntra \! + \! { \Sigmaw^2 } \big ) }{\big (\alpha_m \Ptra \Ntra \! + \! {\Sigmaw^2 } \! + \! \omega_1 { \Sigmap^2 } \big )^2}, \\
	\widehat{ \alpha}^1_m \!=\!  & \,\mathbb{VAR} \{\widehat{\chi}_m | {\cal H}_1 , { \widehat{\cal H}_0}  \} \!=\! \frac{ \alpha_m^2 \Ptra \Ntra \big ( \alpha_m \Ptra \Ntra \! + \! { \Sigmaw^2 } \! + \! {\Sigmap^2 } \big ) }{\big ( \alpha_m  \Ptra \Ntra + { \Sigmaw^2 }  \! + \! \omega_1 { \Sigmap^2 } \big )^2 }. \label{var_H1}
	\end{align}
\end{subequations}
%
%
\noindent Therefore, $\widehat{\alpha}_m = { \omega_0} \,\widehat{\alpha}^0_m + { \omega_1} \,\widehat{\alpha}^1_m$. Also, let $\widetilde{ \alpha}^0_m$ and $\widetilde{ \alpha}^1_m$  indicate the variances of $\widetilde{\chi}_m$ under ${\cal H}_0$ and ${\cal H}_1$, respectively. We have
%
\begin{subequations} \label{ErrVariance}
	\begin{align}
	\widetilde{ \alpha}^0_m = & \,\mathbb{VAR} \{\widetilde{\chi}_m| {\cal H}_0, \widehat{\cal H}_0 \}= \alpha_m - \widehat{\alpha}^0_m,  \\
	\widetilde{ \alpha}^1_m = & \,\mathbb{VAR} \{\widetilde{\chi}_m| {\cal H}_1, \widehat{\cal H}_0  \}= \alpha_m - \widehat{\alpha}^1_m.
	\end{align}
\end{subequations}
%
Hence,  $\widetilde{\alpha}_m = { \omega_0} \,\widetilde{\alpha}^0_m + { \omega_1} \,\widetilde{\alpha}^1_m$. For perfect spectrum sensing, we get $\omega_0 = 1$ and $\omega_1 = 0$ and $\widehat{\chi}_m$ becomes Gaussian.
%
%
\subsection{Determining the Beam Corresponding to \SUrx ~Direction } \label{SubSe3}
\SUrx ~finds $\widehat{\chi}_m$ for all beams Consider the random variable $\widehat{\nu}_m =|\widehat{\chi}_m|^2$. Under hypothesis $ {\cal H}_\ell , \ell = 0 ,1$,  given $\widehat{\cal H}_0$, $\widehat{ \nu}_m$ is an exponential random variable with mean $\widehat{\alpha}^\ell_m$ and pdf 
%
\begin{equation} \label{fvl}
f^\ell_{\widehat{\nu}_m}(y) = \frac{1}{ \widehat{\alpha}^\ell_m} e^{\frac{-y}{ \widehat{\alpha}^\ell_m}  }.
\end{equation}
%
\noindent Hence, the pdf of $ \widehat{\nu}_m$ can be written as 
\begin{equation}
f_{\widehat{\nu}_m }(y) = \omega_0 \,f^0_{\widehat{\nu}_m}(y) + \omega_1 \, f^1_{\widehat{\nu}_m}(y).
\end{equation}
%
\SUrx ~obtains $\widehat{\nu}^*=\max \{\widehat{\nu}_m \}$ among all beams and the corresponding beam index $\mSR^* = {{ \arg\max}} \{ \widehat{\nu}_{m} \}$ and feeds back this information to \SUtx. 
Let $\Psi_i^\ell = \Pr \{\mSR^*=i | {\cal H}_\ell, { \widehat{\cal H}_0} \}$ denote the probability that $\mSR^*=i$ under hypothesis ${\cal H}_\ell$ and the binary detector outcome is $\widehat{\cal H}_0$. To characterize $\Psi_i^\ell$ we need to find the CDF and pdf of $\widehat{ \nu} ^*$ given ${\cal H}_\ell$, denoted as $F^\ell_{\widehat{\nu}^*}(\cdot)$ and $f^\ell_{\widehat{\nu}^*}(\cdot)$, respectively.  Note that given our assumptions, $\widehat{\nu}_m$'s are independent across sectors, however, not necessarily identically distributed. Therefore, the CDF $F^\ell _{\widehat{\nu}^*}(x)$  can be written as
\vspace{-2mm}
\begin{equation}\label{F_sc_l}
 F^\ell_{\widehat{\nu}^* }(y) \!=\! \prod_{m=1}^{M}  F^\ell_{\widehat{\nu}_m} (y ) \!=\! 1 \!+\! \sum_{m=1}^{M} (-1)^{m} \sum\nolimits_{m} e^{ -y A^\ell_{j_1:j_m} }
\end{equation}
%
\vspace{-2mm}
\begin{equation*} 
A^\ell_{j_1:j_m} \! = \! \sum_{i=1}^{m} \frac{1}{    \widehat{\alpha}^\ell_{j_i}    }, \qquad \sum\nolimits_{m} \!\! = \!\! \sum_{j_1=1}^{M-m+1}~ \sum_{j_2=j_1+1}^{M-m+2} \! \cdots \!\! \sum_{j_m =j_{m-1}+1}^{M}\!\!\!.
\vspace{-2mm}
\end{equation*}
%
%
\noindent From the CDF in \eqref{F_sc_l}, we can find the pdf $f^\ell_{\widehat{\nu}^* }(y)$
%
\begin{align}
f^\ell_{\widehat{\nu}^* }(y) = & \sum_{i=1}^{M} f^\ell_{\widehat{\nu}_i} \!(y) \prod_{ \underset {m\neq i}{m=1 }}^{M} \!F^\ell_{\widehat{\nu}_m} \!(y)  \nonumber \\
= & \sum_{m=1}^{M} (-1)^{m+1} \sum\nolimits_{m}  A^\ell_{j_1:j_m} e^{ -y A^\ell_{j_1:j_m} }.
\end{align}
%
Similar to section \ref{PU_Beam_Selection}, we obtain $\Psi_i^\ell$ as

%
\begin{equation}\label{Prob_i}
\Psi_i^\ell =  \int_{0}^{\infty}  f^\ell_{\widehat{\nu}_{i}} (y)  \prod_{\underset{m \neq i} {m=1} }^{M} \! F^\ell_{\widehat{\nu}_{m}} (y) \,dy.
\end{equation}
%
Without loss of generality, suppose $i=1$. After some mathematical simplification, $\Psi_1^\ell$ can be expressed as
%
\begin{equation}
\Psi_1^\ell  = 1+ \sum_{ {m=1} }^{M-1} (-1)^{m} \sum\nolimits_{m}^{\prime} \frac{1}{1 + \widehat{\alpha}^\ell_1   B^\ell_{j_1:j_m} },
\end{equation}
%
where
%
\begin{equation*}
B^\ell_{j_1:j_m} \!\! = \! \sum_{i=1}^{m} \frac{1}{  \widehat{\alpha}^\ell_{1+j_i }   } , \qquad \sum\nolimits_{m}^{\prime} \!\! = \!\! \sum_{j_1=1}^{M-m}~ \sum_{j_2=j_1+1}^{M-m+1} \! \cdots \!\!\! \sum_{j_m =j_{m-1}+1}^{M-1}\!\!\!\!.
\end{equation*}
%
Then, we have $\Psi_i = \Pr \{\mSR^*=i | { \widehat{\cal H}_0} \} = \omega_0 \,\Psi_i^0  +  \omega_1 \,\Psi_i^1$.
%
%
%
\section{Data Transmission Phase}\label{Phase3}
During  this phase, \SUtx ~sends Gaussian data symbols to \SUrx, while data symbol transmission power is adapted based on the  information provided by \SUrx ~through the feedback channel. In particular, \SUtx ~transmits $x(n) \sim {\cal CN} \big (0,P \big )$  over the selected beam $i=\mSR^*$, where $P$ depends on $\widehat{\nu}_i$, and symbols are  independent and identically distributed (i.i.d). Let $u(n)$ denote the discrete-time representation of received signal at \SUrx ~from  $i$-th beam of \SUtx. We note that \SUtx ~enters this phase when the outcome of the binary detector in Section \ref{Eigenvalue_Detector}  is $\widehat{\cal{H}}_0$. Due to error in spatial spectrum sensing, we need to distinguish the signal model for $u(n)$ under ${\cal H}_0$ and ${\cal H}_1$. 
We have
%
%
\begin{equation} \label{DataModel}
\begin{aligned}
{\cal H}_0, \widehat{\cal H}_0: & ~~~u(n) =  \chi_i \,x(n) + { q(n)}, \\
{\cal H}_1, \widehat{\cal H}_0: & ~~~u(n) = \chi_i \,x(n) + { \hsp (n)}  \,s(n) + { q(n)},
\end{aligned}
\end{equation}
%
%
where { $q(n) \sim {\cal CN}(0,\Sigmaq^2)$ and are i.i.d.}
Substituting $\chi_i = \widehat{\chi}_i + \widetilde{ \chi}_i$ in \eqref{DataModel}, we reach at
%
%
\begin{equation} \label{etaEquation}
\begin{aligned}
{\cal H}_0, \widehat{\cal H}_0 : & ~~~u(n) = \widehat{\chi}_i \,x(n) + \overbracket{ \widetilde{ \chi}_i \,x(n) + { q(n)} }^{\text{new noise} ~\eta_{i,0}(n)}, \\
{\cal H}_1, \widehat{\cal H}_0 : & ~~~u(n) = \widehat{\chi}_i \,x(n) + \underbracket{\widetilde{ \chi}_i \,x(n) + { \hsp (n)}  s(n) + { q(n)} }_{\text{new noise} ~\eta_{i,1}(n)}. 
\end{aligned}
\end{equation}
%
%
We obtain an achievable rate expression for a frame by considering symbol-wise mutual information between channel input and output over the duration of  $\Nd$ data symbols as follows
%
%
\begin{align} \label{RFormula}
R & = { \Dd} \sum_{n=1}^{\Nd} \mathbb{E} \left \{I \left ( x(n); u(n) \big |\widehat{\boldsymbol{\nu} }, \widehat{\cal H}_0 \right ) \right \} \nonumber \\
&= { \Dd} \sum_{n=1}^{\Nd} \bigg [ \beta_0 \,\mathbb{E} \left \{ I \left ( x(n); u(n) \big | \widehat{\boldsymbol{\nu}}, {\cal H}_0 ,\widehat{\cal H}_0 \right ) \right \} \nonumber  \\
&~~~~~~~~~~ \,+ \beta_1 \,\mathbb{E} \left \{ I \left (x(n); u(n) \big |\widehat{ \boldsymbol{\nu} }, {\cal H}_1, \widehat{\cal H}_0 \right ) \right \} \bigg ],
\end{align}
%
%
\noindent where $\Dd = \Td/\Tf$ is the fraction of the frame used for data transmission and the expectations are taken over { $\widehat{\boldsymbol{\nu}} = [\widehat{\nu}_1, \ldots, \widehat{\nu}_M]$ } given $\widehat{\cal H}_0$ and ${\cal H}_\ell, \ell=0,1$. To characterize $R$ in \eqref{RFormula} we need to find $\mathbb{E} \big \{ I \big (x(n); u(n) \big | { \widehat{ \boldsymbol{ \nu}} }, {\cal H}_\ell, \widehat{\cal H}_0 \big ) \big \}$ which is given in \eqref{ECGivenH}. Term 1  in \eqref{ECGivenH} is the mutual information between  $x(n)$ and $u(n)$ when \SUtx ~transmits over $j$-th beam, given the estimated channel gain $ \widehat{ \nu}_j = |\widehat{\chi }_j|^2$, and given  ${\cal H}_\ell$ and $\widehat{\cal H}_0$. Term 2 in \eqref{ECGivenH} is the pdf of estimated channel gain $ \widehat{ \nu}_j = |\widehat{\chi }_j|^2$ when $j$-th beam is the selected strongest beam, and is characterized by statistics of channel estimation error and beam selection error, occurred during channel training phase.
%
%
\begin{figure*}
	\begin{align} \label{ECGivenH}
	\mathbb{E} \Big \{ I \left (x(n); u(n) \big | { \widehat{ \boldsymbol{ \nu}} }, {\cal H}_\ell, \widehat{\cal H}_0 \right ) \Big \} = &\int_{\widehat{\nu}_1=0}^{\infty} I \left ( x(n) ;u(n)  \big | \widehat{\nu}_1, \widehat{\cal H}_0, {\cal H}_\ell  \right ) f^\ell_{ \widehat{ \nu}_1} (\widehat{ \nu}_1 )   \Pr \Big (v_1> v_m ~\text{for} ~m=2,...,M|{\cal H}_\ell, \widehat{\cal H}_0 \Big ) \,d\widehat{\nu}_1 \nonumber \\
	+ & \ldots \nonumber \\
	+ &\int_{\widehat{\nu}_M=0}^{\infty} \!\!\!\!\!I \left ( x(n) ;u(n)  \big | \widehat{\nu}_M, \widehat{\cal H}_0, {\cal H}_\ell  \right ) \! f^\ell_{ \widehat{ \nu}_M} (\widehat{ \nu}_M )  \Pr \! \Big (v_M> v_m ~\text{for} ~m=1,...,M\!- \!1|{\cal H}_\ell, \widehat{\cal H}_0 \! \Big ) \,d\widehat{\nu}_M \nonumber \\ 
	= &\sum_{j=1}^{M} \int_{\widehat{\nu}_j=0}^{\infty} \underbracket{ I \left ( x(n) ;u(n)  \big | \widehat{\nu}_j, \widehat{\cal H}_0, {\cal H}_\ell  \right ) }_{\text{Term 1} } \underbracket{ f^\ell_{ \widehat{ \nu}_j} (\widehat{ \nu}_j )  \prod_{\underset{ m \neq j}{m= 1} }^M F^\ell_{ \widehat{ \nu}_m} ({ \widehat{ \nu}_j}) }_{ \text{Term 2} } \,d\widehat{\nu}_j.
	\end{align}
	\hrulefill
\end{figure*}
%
\noindent Focusing on Term 1 in \eqref{ECGivenH} we have
%
\begin{align} \label{DEntropy}
I \left (x(n);u(n) \big | \widehat{\nu}_i, \widehat{\cal H}_0, {\cal H}_\ell  \right ) = &\,h \big ( x(n) \big | \widehat{\nu}_i, \widehat{\cal H}_0, {\cal H}_\ell \big ) \nonumber \\
- &\,h \big ( x(n) \big | u(n), \widehat{\nu}_i, \widehat{\cal H}_0, {\cal H}_\ell \big ),
\end{align}
%
where $h(\cdot)$ is the differential entropy. From now on, we drop the variable $n$ in $x(n)$ and $u(n)$ for brevity. Consider the first term in \eqref{DEntropy}. Since $x \sim {\cal CN} (0,P)$ we have $ h \big ( x| \widehat{\nu}_i, \widehat{\cal H}_0, {\cal H}_\ell \big )= \log_2( \pi e P)$. Consider the second term in \eqref{DEntropy}. 
Due to channel estimation error,  the new noises $\eta_{i,\ell}$ in \eqref{etaEquation} are non-Gaussian  and this term does not have a closed form expression. Hence, similar to \cite{Hassibi1,  Azadeh2, Vosoughi_Capacity}  we employ bounding techniques to find an upper bound on this term. This term is upper bounded by the entropy of a Gaussian random variable with the variance $\Theta_{\rm {M}}^{i, \ell}$ 
%
\begin{equation}
\Theta_{\rm {M}}^{i, \ell} = \mathbb{E} \left \{ \big | x-\mathbb{E} \big \{ x \,| \,\widehat{\nu}_i, \widehat{\cal H}_0, {\cal H}_\ell \big \}  \big |^2 \right \},
\end{equation}
%
%
where the expectations are taken over the conditional pdf of $x$ given $u,\widehat{\nu}_i, \widehat{\cal H}_0, {\cal H}_\ell$.  In fact, $\Theta_{\rm {M}}^{i, \ell}$ is the mean square error (MSE) of the MMSE estimate of $x$ given $u,\widehat{\nu}_i, \widehat{\cal H}_0, {\cal H}_\ell$. Using minimum variance property of MMSE estimator, we have  $\Theta_{\rm {M}}^{i, \ell} \leq \Theta_{\rm {L}}^{i, \ell}$, where $\Theta_{\rm {L}}^{i, \ell}$ is the MSE of the LMMSE estimate of $x$ given $u,\widehat{\nu}_i, \widehat{\cal H}_0, {\cal H}_\ell$.  Combining all, we find $h(x |u, \widehat{\nu}_i, \widehat{\cal H}_0, {\cal H}_\ell) \leq \log_2(\pi e \Theta_{\rm {L}}^{i, \ell})$ and $I(x,u | \widehat{\nu}_i, \widehat{\cal H}_0,{\cal H}_\ell) \geq \log_2 (P/\Theta_{\rm {L}}^{i, \ell})$ where
%
%
\begin{equation}
\Theta_{\rm {L}}^{i, \ell} =  \frac{P \sigma_{\eta_{i,\ell}}^2 }{ \sigma_{\eta_{i,\ell}}^2 + \widehat{\nu}_i P}, \qquad
\sigma_{\eta_{i,\ell}}^2 =  \widetilde{\alpha}^\ell_i P + \Sigmaq^2 + \ell \Sigmap^2.
\end{equation}
%
At the end, we obtain the lower bounds as follow
%
\begin{subequations}\label{IH0H1}
	\begin{align}
	\!\!\! I \left (x(n);u(n) \big | \widehat{\nu}_i, \widehat{\cal H}_0, {\cal H}_0  \right )  \geq &\log_2 \! \Big(\!  1 \! + \! \frac{ {\widehat{\nu}_i } P }{ \widetilde{ \alpha}^0_i P \! +  \! \Sigmaq^2}  \Big), \\
	\!\!\! I \left (x(n);u(n) \big | \widehat{\nu}_i, \widehat{\cal H}_0, {\cal H}_1  \right ) \geq &\log_2\! \Big(\! 1 \! + \! \frac{ \widehat{\nu}_i P }{\widetilde{ \alpha}^1_i P \! + \! \Sigmaq^2 \! + \! \Sigmap^2  }  \Big).
	\end{align}
\end{subequations}
%
%
Substituting equations \eqref{ECGivenH} and \eqref{IH0H1} in \eqref{RFormula} and changing the integration variable (replacing $\widehat{\nu}_j$ with $y$), we reach at
%
%
\begin{align}\label{C_Ergodic}
R \! \geq \! \RLB \! =& { \Dd} \beta_0 \!\! \sum_{j=1}^{M} \!\! \int \!\! \log_2 \! \Big( \! 1\!+\!\frac{ \widehat{\nu}_j P }{\widetilde{ \alpha}^0_j P \! + \! \Sigmaq^2} \! \Big ) \! f^0_{\widehat{\nu}_j} \! (\widehat{\nu}_j) \!\!\prod_{ \underset{m \ne j} {m=1}}^M \!\!\!F^0_{ \widehat{\nu}_m} \!(\widehat{\nu}_j) d\widehat{\nu}_j \nonumber\\
+  \,{ \Dd} \beta_1 &\!\! \sum_{j=1}^{M} \!\!\int \!\!\log_2 \! \Big( \! 1\!+\!\frac{ \widehat{\nu}_j P }{\widetilde{ \alpha}^1_j P \! +\! \Sigmaq^2 \! +\! \Sigmap^2} \! \Big ) \! f^1_{ \widehat{\nu}_j} (\widehat{\nu}_j)\!\!\prod_{ \underset{m \ne j} {m=1}}^M \!\!\!F^1_{\widehat{\nu}_m} \!(\widehat{\nu}_j) d\widehat{\nu}_j.
\end{align}
%
%
\noindent We note that the lower bounds in \eqref{IH0H1} are achieved when the new noises $\eta_{m,0},\eta_{m,1}$ in \eqref{etaEquation} are regarded as worst-case Gaussian noise and hence the MMSE and LMMSE of $x$ given $u, \widehat{\nu}_m, \widehat{\cal H}_0, {\cal H}_\ell$ coincide.
\par So far, we have established a lower bound on the achievable rates. Next, we characterize AIC and ATPC. Let $\Ibar$ indicate the maximum allowed interference power imposed on PU. To satisfy the AIC, we need to have 
\vspace{-2mm}
%
\begin{align}\label{Iav0}
\beta_1 \,\mathbb{E}  \big \{ g\big \} \Bigg [& \,{ \Dd} \mathbb{E}  \big\{p( \kappa_{\rm{SR} }^* \! -  \! \kappa_{ \rm{PU}}^*) \,P  \,| \,  {\cal H}_1, { \widehat{\cal H}_0} \big \}  \\ 
& { \,+ \Dtr \Ptra  \sum_{j=1}^{M} \mathbb{E}  \big\{ p( \kappa_{j } \! - \!  \kappa_{\rm{PU}}^*) \,| \,  {\cal H}_1, { \widehat{\cal H}_0} \big \} } \Bigg ] \leq \Ibar, \nonumber
\end{align}
%
where $\Dtr = \Ttra/\Tf$. The first term in \eqref{Iav0} is the average interference imposed on PU when \SUtx ~transmits data symbols, and the second term  is the average interference imposed on PU when \SUtx ~sends training symbols for channel estimation at \SUrx. Consider the two conditional expectation terms inside the bracket in \eqref{Iav0}. Using the fact that, given ${\cal H}_1, { \widehat{\cal H}_0}$, $p(\cdot)$ and $P$ (which depends on $\widehat{\nu}^*$) are independent, and also the average probabilities derived  in \eqref{DeltaAverage} and \eqref{Prob_i} we have 
%
\begin{equation}\label{ExpecP} 
\mathbb{E} \big \{ p ( \kappa_{ \rm{SR}}^*  - \kappa_{\rm{PU}}^* ) | {\cal H}_1, { \widehat{\cal H}_0}  \big \}  = \sum_{j=1}^{M} \sum_{i=1}^{M} \Psi^1_ j ~\overline{\Delta}_{\mPU^*, i} ~ p ( \kappa_{j}  - \kappa_{i} ), 
\end{equation}
%
\vspace{-2mm}
\begin{equation}
\mathbb{E}  \big\{ p( \kappa_{j } - \kappa_{\rm{PU}}^*) \,| \,  {\cal H}_1, { \widehat{\cal H}_0} \big \} = \sum_{i=1}^{M} \overline{\Delta}_{\mPU^*, i} ~ p ( \kappa_{j}  -  \kappa_{i} ).
\end{equation}
%
Then, the constraint in \eqref{Iav0} can be written as 
%
\begin{equation}\label{Iav}
{ \Dd} \,b_0 \,\mathbb{E} \big \{ P | {\cal H}_1, { \widehat{\cal H}_0} \big \} +  \Dtr u_0 \Ptra  \leq \Ibar, 
\end{equation}
\vspace{-1mm}
%
where
\vspace{-2mm}
\begin{subequations}
	\begin{align}
	b_0 = & \,\beta_1 { \gamma} \sum_{j=1}^{M} \sum_{i=1}^{M} \Psi_j^1  ~\overline{\Delta}_{\mPU^*, i} ~ p ( \kappa_{j}  - \kappa_{i} ), \label{b0}\\
	u_0 =  & \,\beta_1 { \gamma}  \sum_{j=1}^{M} \sum_{i=1}^{M} \overline{\Delta}_{\mPU^*, i} ~ p ( \kappa_{j}  - \kappa_{i} ), \label{u0} \\
	\mathbb{E} \big \{ &P | {\cal H}_1, { \widehat{\cal H}_0} \big \} =\int P(y) f^1_{\widehat{\nu}^*} (y) dy.
	\end{align}
\end{subequations}
%
%
\noindent We note that spectrum sensing error, PU beam selection error, and \SUrx ~beam selection error are reflected in AIC through variables $\beta_1$, $\overline{\Delta}_{\mPU^*, i}$ and $\Psi_j^1$, respectively. Also, channel estimation error influences AIC through variable $P$.  Let $\Pbar$ denote the maximum allowed average transmit power of \SUtx. To satisfy  the ATPC, we need to have
%
\begin{equation}\label{Pav0}
\beta_0 { \Dd} \mathbb{E} \big \{ P  | {\cal H}_0, { \widehat{\cal H}_0} \big \} \! + \! \beta_1 { \Dd} \mathbb{E} \big \{ P  |  {\cal H}_1, { \widehat{\cal H}_0}  \big \}  \! + \! \pioh \Dtr \Ptra \leq \Pbar,
\end{equation}
%
%
\noindent where $	\mathbb{E} \{ P | {\cal H}_0, { \widehat{\cal H}_0} \} =\int P(y) f^0_{\widehat{\nu}^*} (y) dy$, and  the third term in \eqref{Pav0} accounts for transmit power used for training symbols. We note that spectrum sensing error affects ATPC through variables $\beta_0$, $\beta_1$ and $\pioh$. Also, channel estimation error affects ATPC through variable $P$. 
\par Now that we have characterized a lower bound on the achievable rates $\RLB$ in \eqref{C_Ergodic}, AIC in \eqref{Iav}, and ATPC in \eqref{Pav0}, we summarize how the four error types, namely, spectrum sensing error, beam detection error, channel estimation error, and beam selection error, affect these expressions. {First}, spectrum sensing error affects AIC via $\beta_1$, both ATPC and $\RLB$ via $\beta_0$ and $\beta_1$. Recall $\beta_0,\beta_1$ depend on $\pi_0, \overline{P}_{\rm{fa}}, \overline{P}_{\rm{d}} $ (see \eqref{Beta}). {Second,} beam detection error affects AIC via $\overline{\Delta}_{\mPU^*, i}$ and does not have a direct impact on ATPC and $\RLB$. {Third}, channel estimation error affects both AIC and ATPC via {$\Ttra$}, and  $\RLB$ via $\widetilde{\alpha}_m^\ell$. {Fourth}, beam selection error impacts AIC, ATPC and $\RLB$ via $P$ (which depends on the estimation channel gain of the selected beam). 

\par Having the mathematical expressions for $\RLB$, AIC, ATPC, our goal is to allocate transmission resources such that $\RLB$ is maximized, subject to the aforementioned constraints. To determine our optimization variables, we need to examine closely the underlying trade-offs between decreasing average interference and average transmit powers, decreasing four types of errors (i.e., spectrum sensing error,  beam detection error, channel estimation error, and beam selection error), and increasing $\RLB$. Within a frame with fixed duration of $\Tf$ seconds, time is divided between three phases with variable durations: spatial spectrum sensing with duration $\Tsen$, channel training with duration $\Ttra$, and data transmission with duration of $\Td$. Suppose $\Tsen$ increases. On the positive side, spectrum sensing error, beam detection error, and average interference imposed on PU decrease (i.e., for ideal spectrum sensing $\beta_1=0$ in \eqref{Beta} and data transmission from \SUtx ~to \SUrx ~does not cause interference on PU). On the negative side, $\Ttra + \Td$ decreases, that can lead to increasing channel estimation error (due to decrease in $\Ttra$) and/or decreasing $\RLB$ (due to decrease in $\Td$).  Given $\Tsen$, as $\Ttra$ increases, channel estimation error in \eqref{ErrVariance} decreases. However, average interference imposed on PU during transmission of training symbols increases and $\RLB$ decreases \footnote{Note that as channel estimation error in \eqref{ErrVariance} decreases, the lower bounds in \eqref{IH0H1} increase. However, this logarithmic increase is dominated by the linear decrease of $\Dd$ in \eqref{C_Ergodic}, which leads into a decrease in $\RLB$.}. Finally, increasing data symbol transmission power $P$ increases $\RLB$, however, it increases average interference and average transmit power. Based on all these existing trade-offs, we seek the optimal $\Tsen, \Ttra, P$ such that $\RLB$ in \eqref{C_Ergodic} is maximized, subject to AIC and ATPC given in \eqref{Iav} and \eqref{Pav0}, respectively. In other words, we are interested in solving the following constrained optimization problem
%
%
\leqnomode
\begin{align*}\tag{P1}\label{Prob1}
\hspace{-5mm} { \underset { \Tsen, \Ttra, P  }  {\text{Maximize}} }  ~~\RLB \\
\end{align*}
\vspace{-12mm}
\begin{align}
\begin{array}{lll}
\text{s.t.:}  ~~& 0  < \Tsen  < \Tf - \Ttra \nonumber \\
 & \Ttra >0 , \quad P \geq 0 \nonumber \\
 & \eqref{Iav}  ~\text{and} ~ \eqref{Pav0} ~\text{are satisfied.} \nonumber
 \end{array}
 \vspace{0mm}
\end{align}
\reqnomode
%
%
%
\section{Constrained Maximization of Rate Lower Bound} \label{Solving}
In this section, we address the optimization problem \eqref{Prob1}. Taking the second derivative of $\RLB$ with respect to (w.r.t.) the optimization variables, we note that \eqref{Prob1} is not jointly concave over $\Tsen$, $\Ttra$, $P$. However, given $\Tsen$ and $\Ttra$, \eqref{Prob1} is concave w.r.t. $P$\footnote{The cost function of \eqref{Prob1} given in \eqref{C_Ergodic} depends on $P$ through the two logarithms, that can be viewed, in terms of $P$, as $(1+ \frac{ aP}{bP+c})$, where $a, b,c$ are positive. Since the arguments of these logarithms  are concave, $\RLB$ is also concave w.r.t. $P$.}. We propose an iterative method based on the block coordinate descent (BCD) algorithm to solve \eqref{Prob1}. The underlying principle of the BCD algorithm is that, at each iteration one variable is optimized, while the remaining variables are fixed. The iteration continues until it converges to a stationary point of \eqref{Prob1} \cite{Moji1}. To apply the principle of the BCD algorithm to \eqref{Prob1}, we consider the following three steps. \textbf {Step (i)}: given $\Tsen$, $\Ttra$, we optimize $P$ using the Lagrangian method. The Lagrangian is
%
%
\begin{equation}\label{Lagran}
{\cal L} =  - R_\text{LB} + \mu \big [ \text{LHS of} ~\eqref{Iav} - \Ibar \big ]  
+ \lambda \big [ \text{LHS of} ~\eqref{Pav0} - \Pbar \big],
\end{equation}
%
%
\noindent in which LHS stands for left-hand side, $\lambda$ and $\mu$ are the nonnegative Lagrange multipliers, associated with the  ATPC and AIC, respectively. Therefore, the optimal $P$ that minimize \eqref{Lagran} is the solution to the Karush-Kuhn-Tucker (KKT) optimality necessary and sufficient conditions. The KKT conditions are the first derivatives of ${\cal L}$ w.r.t. $P, \mu, \lambda$ being equal to zero, i.e., $\partial {\cal L}/\partial P = 0, \partial {\cal L}/\partial \mu = 0, \partial {\cal L}/\partial \lambda = 0$. We have
%
%
\begin{subequations} \label{Rond_Lag}
	\begin{align}
	- {1 \over \ln(2) } &\sum_{\ell=0}^{1} {\beta_\ell} \sum_{i=1}^{M}   \frac{y \, (\Sigmaq^2 + \ell \Sigmap^2 ) \,f^\ell_{ \widehat{\nu}_i } (y) }{ \sigma_{\eta_{i,\ell}}^2 \big ( y P +  \sigma_{\eta_{i,\ell}}^2 \big )} \prod_{ \underset{m\neq i}{ m =1}}^{M} F^\ell_{\widehat{\nu}_m } (y) \nonumber \\
	+ & \lambda \Big [  \beta_0  f^0_{\widehat{\nu}^* }(y)  +  \beta_1  f^1_{\widehat{\nu }^* }(y) \Big ] + \mu  b_0  f^1_{ \widehat{\nu }^* } (y)  = 0, \label{Rond_Lag_P}\\
	&\mu \,\Big | \text{LHS of} ~\eqref{Iav} - \Ibar \big |  = 0,  \label{Rond_Lag_mu}\\
	&\lambda  \,\Big | \text{LHS of} ~\eqref{Pav0} -  \Pbar \Big |  =   0. \label{Rond_Lag_lam}
	\end{align}
\end{subequations}
%
The closed-form analytical solution for \eqref{Rond_Lag} cannot be found. Hence,  we solve these equations numerically for every realization of $\widehat{\nu}^*$, via the following iterative method. We first initialize the Lagrangian multipliers  $\mu$ and $\lambda$ and then find $P$ using \eqref{Rond_Lag_P}. Next, we update $\mu$ and $\lambda$ using the subgradient method. Using the updated $\mu$ and $\lambda$, we find $P$ again using \eqref{Rond_Lag_P}. We repeat this procedure until $\mu$ and $\lambda$ converge (i.e.,
a pre-determined stopping criterion is met). \textbf {Step (ii)}: given  $P$ and $\Ttra$, we optimize $\Tsen$. We note that the optimal $\Tsen$ lies in the interval $(0, \Tf-\Ttra)$. We find it using numerical search methods (e.g., bisection method). \textbf{Step (iii)}: given $P$ and $\Tsen$, we optimize $\Ttra$. We note that the optimal $\Ttra$ lies in the interval $(0, \Tf-\Tsen)$ and we find it using numerical search methods.
%
%
\begin{figure}[!t]
	\centering
	\includegraphics[width=55mm]{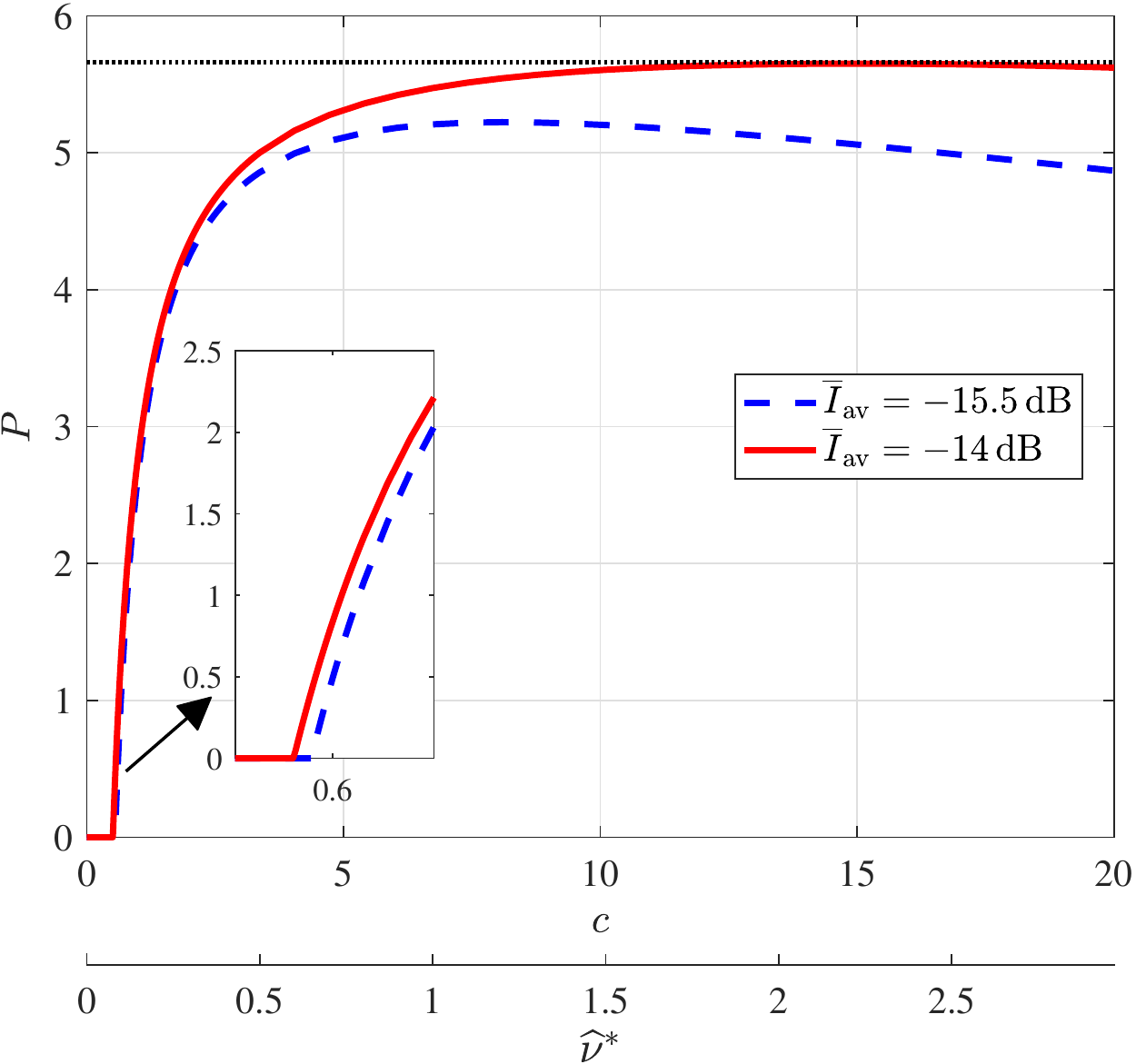}
	\caption{The optimized $P$ obtained from \eqref{Rond_Lag_P} versus $\widehat{\nu}^*$  (and $c$) for $\Pbar=2\,$dB.} 
	\label{PvsG}      
\end{figure}
%
\par To gain an insight on the solution of \eqref{Prob1}, we look into the behavior of the optimized $P$ versus the realizations of the estimated channel gain $\widehat{\nu}^*$. Fig. \ref{PvsG} illustrates the optimized  $P$ versus $\widehat{\nu}^*$ (and $c$, where $\widehat{\nu}^* = c \, m_{ \widehat{ \nu}^* }$ and  $m_{ \widehat{ \nu}^* }$ is the mean of $\widehat{ \nu}^*$) for $\Ibar = -15.5, -14\,$dB and other simulation parameters given in Table \ref{table2}. For these parameters { 
$m_{ \widehat{ \nu}^* } = 0.1484$}. We observe that the optimized $P$ for very small $\widehat{\nu}^*$ (when $\widehat{\nu}^*$ is smaller than a cut-off threshold $\zeta = 3.5 \,m_{ \widehat{ \nu}^* }$) is zero.  As $\widehat{\nu}^*$ increases the optimized $P$ increases gradually until it reaches a maximum value. As $\widehat{\nu}^*$ increases further, the optimized $P$ decreases, until it reaches a minimum value for very large $\widehat{\nu}^*$ (when $\widehat{\nu}^* > 85 \,m_{ \widehat{ \nu}^* }$), not shown in the figure. Comparing the curves for $\Ibar=-15.5\,$dB and $\Ibar=-14\,$dB, we note that the optimized $P$ decays faster (after it reaches its maximum value) for lower $\Ibar$. Moreover, the cut-off threshold  $\zeta$ is lower for higher $\Ibar$. The behavior of the optimized $P$ versus $\widehat{\nu}^*$ is different from our intuitive expectation that expects to see the optimized $P$ increases monotonically as $\widehat{\nu}^*$ increases. We explore this by examining the optimized $P$,  which satisfies \eqref{Rond_Lag_P}.
\par Although for general $M$ the optimized  $P$ does not have a closed form expression, for $M=1$  and under a simplifying assumption\footnote{We assume that the optimized $\Ttra$ is large enough such that $\widetilde{ \alpha} P + \Sigmaq^2 \approx \Sigmaq^2$. This assumption allows us to approximate \eqref{Rond_Lag_P} for $M=1$ as a quadratic polynomial in $P$ (originally a polynomial of degree $4$ in $P$) and find a closed-form expression for $P$.} it can approximated as follows:
%
\begin{align} \label{Aprrox_P}
&\qquad \qquad \quad~~ P  \approx  \left [ \frac{F + \sqrt{\Upsilon}}{2}  \right ] ^{+}, \\
F & = \frac{\beta_0 W(\widehat{\nu}^*) \! + \!\beta_1 }{\ln(2) \Big [\lambda \big (\beta_0 W(\widehat{\nu}^*) \! +\! \beta_1 \big ) \! + \! \mu b_0  \Big ]} -  \frac{2 \Sigmaq^2 \! + \! \Sigmap^2 }{ \widehat{\nu}^*}, \nonumber
\end{align}
%
\begin{align} 
\Upsilon \! = \! F^2  & \! - \! \frac{4}{\widehat{\nu}^*} \!\! \left ( \!\!\! \frac{ \Sigmaq^2 (\Sigmaq^2 \! + \! \Sigmap^2 )}{ \widehat{ \nu}^*} \! - \! \frac{ \big (\beta_0 W(\widehat{\nu}^*) \!+ \!\beta_1 \big ) \Sigmaq^2 \! + \! \beta_1 \Sigmap^2 }{ \ln(2) \Big [ \lambda (\beta_0 W(\widehat{\nu}^*) \! + \!\beta_1 ) \! + \! \mu b_0  \Big ]} \! \right ) \nonumber \!.
\end{align}
%
\noindent where  $W(\widehat{\nu}^*)=f^0_{ \widehat{\nu}^*}(\widehat{\nu}^*) /f^1_{ \widehat{ \nu}^*}(\widehat{\nu}^*) = \widehat{ \alpha}^1 / \widehat{ \alpha}^0 ~e^{ -\widehat{\nu}^* (\frac{1}{\widehat{ \alpha}^0 } - \frac{1}{ \widehat{ \alpha}^1 } ) }$. Considering \eqref{alpha_hat} we realize that $\widehat{ \alpha}^0 < \widehat{ \alpha}^1$. This implies as $\widehat{\nu}^*$ increases,  $W(\widehat{\nu}^*)$ and $\Upsilon$ decrease. However, the behavior of $F$ changes, i.e., $F$ increases until it reaches a maximum value. As $\widehat{\nu}^*$ increases further, $F$ decreases. Considering \eqref{Aprrox_P} we note that the behavior of $P$ (in terms of $\widehat{\nu}^*$) is dominated by the behavior of $F$. In the ideal scenario when there is no channel estimation error, we have $\widehat{ \alpha}^0 = \widehat{ \alpha}^1 = \alpha$  and $W(\widehat{\nu}^*)=1$, $F$ monotonically increases and  $\Upsilon$ deceases, i.e., $P$ in \eqref{Aprrox_P} monotonically increases as $\widehat{\nu}^*$ increases, which is what we intuitively expect. 
%
%
%
%
\begin{table}[t!]
	\begin{center}
		\caption{$\Pr(\widehat{\nu}^* \geq c \, m_{ \widehat{\nu}^* })$ in terms of $c$, given $m_{ \widehat{\nu}^* } = 0.1484$.}
		\label{table3}
		\begin{footnotesize}
			\begin{tabular}{|l|l|} 
				\hline
			    $c$ & $\Pr(\widehat{\nu}^* \geq c \, m_{ \widehat{\nu}^* })$\\
				\hline \hline
				$4$ & $ 3.01 \times 10^{-3}$ \\ 
				$8$ & $ 7.04 \times 10^{-6}$ \\ 
				$12$ & $ 1.54 \times 10^{-8}$ \\
				$16$ & $ 4.87 \times 10^{-11}$ \\
				\hline
			\end{tabular}
		\end{footnotesize}   
	\end{center}    
	\vspace{0mm}
\end{table}
%
%
\par The optimized $P$ we discussed so far requires solving \eqref{Rond_Lag} several times for each realization of $\widehat{\nu}^*$. Integrating the insights we have gained into how this optimized $P$ varies in terms of $\widehat{\nu}^*$,  we propose two transmit power control schemes that are simpler to implement and yield achievable rate lower bounds that are very close to the maximized $\RLB$ values in \eqref{Prob1}. Since $\Pr( \widehat{\nu}^* \geq c \, m_{ \widehat{ \nu}^*})$ is very small for $c \geq 8$  (see Table \ref{table3}),  we focus on the regime when $\widehat{\nu}^* <  8 \, m_{ \widehat{\nu}^* }$ and develop  two schemes, dubbed here scheme 1 and scheme 2,  that mimic the behavior of the optimized $P$ in this regime. 
%
%
\subsection{Scheme 1}\label{Omni-CPT}
For scheme 1, when the spectrum is sensed idle, \SUtx ~sends data to \SUrx ~over the selected sector $i=\mSR^*$ according to the following rule:
%
\begin{equation}\label{PORA}
P_{\rm{S}_1} = \left\{
\begin{array}{ll} 
{\Pi}_1 ,  & \text{if}~~  \widehat{\nu}^* \geq \zeta_1  \\
0, & \text{if}~~  \widehat{\nu}^* < \zeta_1 \\
\end{array}\right.  
\end{equation}
%
\noindent i.e., when $\widehat{\nu}^*$ is less than a cut-off threshold $\zeta_1$, \SUtx ~remains silent, when $\widehat{\nu}^*$ is larger than $\zeta_1$, \SUtx ~lets its transmit power be equal to constant $\Pi_1$. The parameter $\Pi_1$ can be found in terms of $\Tsen, \Ttra, \zeta_1$, via enforcing AIC in \eqref{Iav} and ATPC in \eqref{Pav0} as the following:
%
\begin{equation}\label{Pi_ora}
\Pi_1 = { 1 \over \Dd} \min \!\left \{ \frac{\Pbar - { \pioh \Dtr \Ptra }  }{ \sum_{\ell =0}^{1} \beta_\ell \big (1\!-\! F^\ell_{ \widehat{\nu}^* } (\zeta_1) \big ) } \, , \, \frac{\Ibar { - u_0 \Dtr \Ptra } }{ b_0 \big (1\!-\!F^1_{ \widehat{\nu}^* }  (\zeta_1)  \big )} \right \}\!.
\end{equation}
%
%
Let $\Rora$ denote the lower bound on the achievable rates when \SUtx ~adopts the power control scheme in \eqref{PORA}. We find  $\Rora$ expression  by substituting $P_{\rm{S}_1}$ in \eqref{C_Ergodic} and taking expectation w.r.t. $\widehat{\nu}^*$. This expression is given in \eqref{CLB_ORA}
%
%
where $\text{SNR}^0_{i} = {\Pi_1 \over \widehat{ \alpha}^0_i + \Sigmaq^2}$,  $\text{SNR}^1_{i} = {\Pi_1 \over \widehat{ \alpha}^1_i + \Sigmaq^2 + \Sigmap^2 }$ and $\text{Ei}(\cdot)$ is the exponential integral.
%
%
\begin{figure*}
	\begin{equation}\label{CLB_ORA}
	\Rora = \,{\Dd \over \ln(2)} \sum_{\ell = 0}^1 \beta_\ell \sum_{j=1 }^{M} \bigg [ Y \big ( \widehat{\alpha }^\ell_j , \text{SNR}_j^\ell \big ) + \sum_{ \underset{m \neq j}{m =1} }^{M} (-1)^{m} \sum\nolimits_{m} Y \big ( d^\ell_{j,m}, \text{SNR}_j^\ell \big ) \bigg ]
	\end{equation}
	\begin{equation*}
	Y(a,b) \! = \!\int_{\zeta_1}^{\infty} \!\!\ln(1+b x) {1 \over a} e^{- x \over a} dx =   e^{-\zeta_1/a} \ln \big (1+b \zeta_1 \big)  -  e^{1/ab}  ~\text{Ei} \big (- \zeta_1/a -1/ab \big), \qquad ~~d^{\ell}_{j, m} \! = \! \left ( A^\ell_{k_1:k_m} + {1 \over \widetilde{ \alpha }_j^\ell } \right )^{-1}.
	\end{equation*}		
	\hrulefill
\end{figure*}
%
%
\noindent With this transmit power scheme, we consider a modified problem to \eqref{Prob1}, where the lower bound $\Rora$ in \eqref{CLB_ORA} is maximized (subject to the same constraints) and the optimization variables are $\Tsen, \Ttra, \zeta_1$. In Section \ref{SimResults} we numerically compare the maximized $\RLB$ in \eqref{Prob1} and the maximized $\Rora$. 
%
%
\subsection{Scheme 2}
For scheme 2, when the spectrum is sensed idle, \SUtx ~sends data symbols to \SUrx ~over the selected sector $i=\mSR^*$ according to the following rule:
%
\begin{equation} \label{PNEW}
P_{\rm{S}_2} = \left\{
\begin{array}{ll} 
\Pi_2 \big (1- \frac{\zeta_2 }{ \widehat{\nu}^* } \big) , & \text{if}~~  \widehat{\nu}^*  \geq \zeta_2 \\
0, & \text{if}~~  \widehat{\nu}^*  < \zeta_2 \\
\end{array}\right. 
\end{equation}
%
Different from scheme 1, in the scheme 2 when $\widehat{\nu}^*$ exceeds the cut-off threshold $\zeta_2$, \SUtx ~transmits at a variable power. The power level increases as $\widehat{\nu}^*$ increases, until it reaches its maximum value of $\Pi_2$,  i.e., $ \lim_{\widehat{\nu}^* \to \infty} P_{\rm{S}_2} = \Pi_2$. The parameter $\Pi_2$ can be found in terms of $\Tsen, \Ttra, \zeta_2$, via enforcing AIC in \eqref{Iav} and ATPC in \eqref{Pav0} as the following:
%
\begin{equation}\label{OmegaNew}
\Pi_2 = { 1 \over \Dd} \min \!\left \{ \frac{\Pbar - { \pioh \Dtr \Ptra } }{ \sum\nolimits_{\ell =0}^{1} \beta_\ell \big [1\!-\! G^\ell(\zeta_2) \big ]  } \, , \, \frac{\Ibar { - u_0 \Dtr \Ptra} }{ b_0 \big [1\!-\!G^1 (\zeta_2) \big ]} \right \},
\end{equation}
%
%
where $G^\ell(\zeta_2) = F^\ell_{ \widehat{\nu}^* } (\zeta_2) + \zeta_2 T^\ell (\zeta_2)$ and
%
%
\begin{align}
T^\ell (\zeta_2)  =  & \,\mathbb{E} \left \{ {1 \over \widehat{\nu}^* } \,\big| \,\widehat{\nu}^* \geq \zeta_2 , \,{\cal H}_\ell \right \} = \int_{\zeta_2 }^{\infty} { f^\ell_{\widehat{\nu}^* }(y) \over y} dy \nonumber  \\
= & \,  \sum_{m=1}^{M} \!(-1)^{m}  \sum\nolimits_{m} A^\ell_{j_1:j_m} \text{Ei}\big (\! - \! \zeta_2 A^\ell_{j_1:j_m}  \big).
\end{align}
%
%
Let $\Rnew$ represent the lower bound on the achievable rates when \SUtx ~adopts the power control scheme in \eqref{PNEW}. We find  $\Rnew$  by substituting $P_{\rm{S}_2}$ in \eqref{C_Ergodic} and taking expectation w.r.t. $\widehat{\nu}^*$. With this transmit power scheme, we consider a modified problem to \eqref{Prob1}, where the lower bound $\Rnew$ is maximized (subject to the same constraints) and the optimization variables are $\Tsen, \Ttra, \zeta_2$. In Section \ref{SimResults} we numerically compare the maximized $\RLB$ in \eqref{Prob1} and the maximized $\Rnew$. Note that the closed-form expression for $\Rnew$ cannot be obtained.
%
%
\section{Simulation Results}\label{SimResults}
\begin{figure}[!t]
	\centering
	\includegraphics[width=55mm]{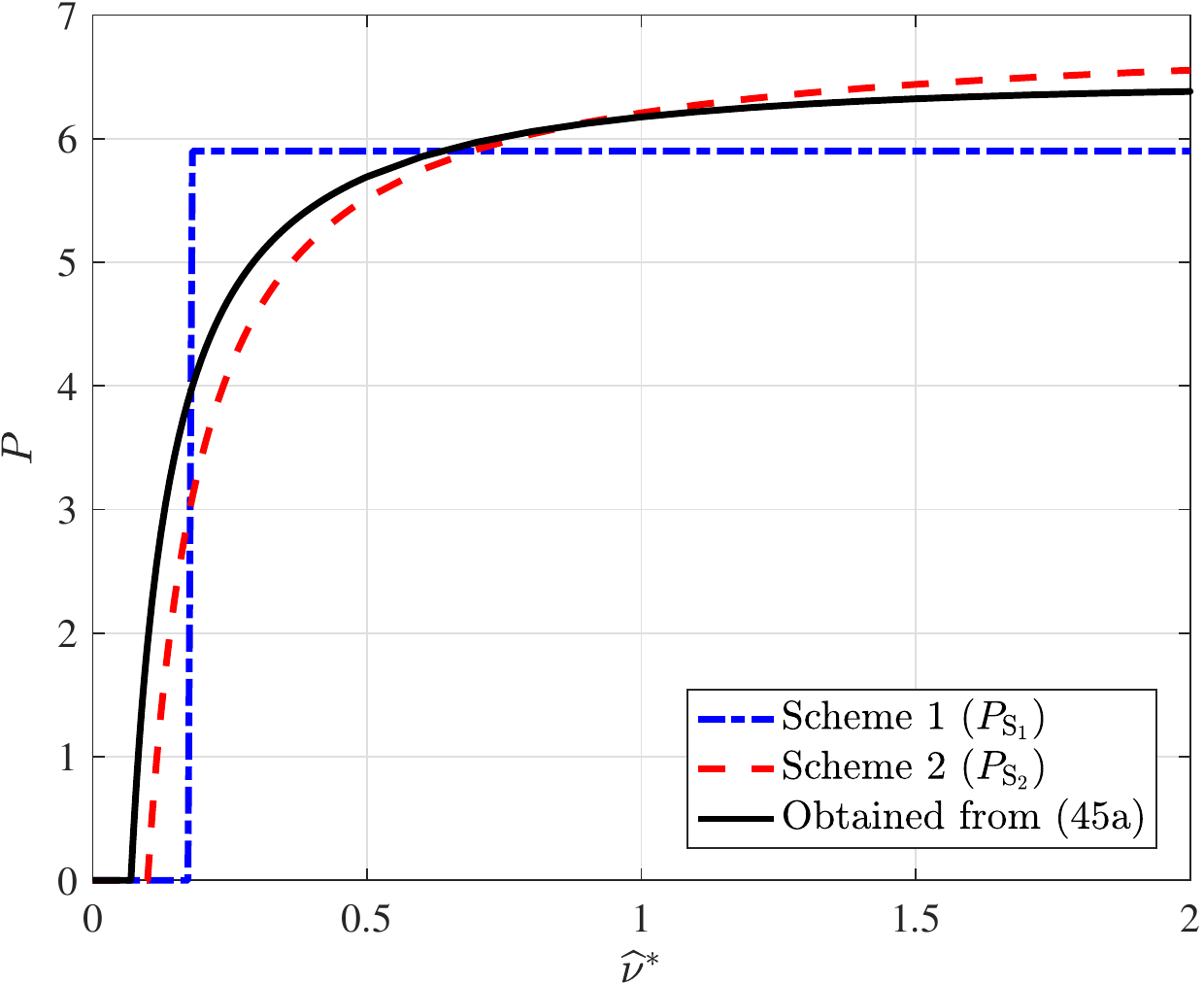}
	\caption{ $P$ versus $\widehat{\nu}^*$ for $\Pbar=2\,\text{dB}, \Ibar=-12\,$dB .} 
	\label{Pv_vs_v}      
\end{figure}

%
%
%
\begin{table}[t!]
	\begin{center}
		\vspace{0mm}
		\caption{Simulation Parameters}
		\label{table2}
		\begin{footnotesize}
			\begin{tabular}{|l|l||l|l||l|l|} 
				\hline
				\textbf{Parameter} & \textbf{Value} & \textbf{Parameter} & \textbf{Value} & \textbf{Parameter} & \textbf{Value}\\
				\hline \hline
				$A_0$ & $0.98$ & $\gamma_{\rm{ss}}$ & $0.1$ & $\Sigmaw^2, \Sigmaq^2$ & $0.5$ \\
				$A_1$ & $0.02$ & $\gamma, \gamma_{\rm{sp}}$ & $0.5$ & $\Pp$ & $0.5$ watts \\
				$\phi_{\rm{3dB}}$ & $20^\degree$ & $\pi_1$ & $0.7$ & $\Tf$ & $30$ ms\\
				$\phi_1$ & $-55^\degree$ & ${\overline{P}}_{\rm{d}}$ & $0.85$ & $\Ptra$ & $2$ watts \\
				$\phi_2$ & $+55^\degree$ & $M$ & $7$ &  & \\
				\hline
			\end{tabular}
		\end{footnotesize}   
	\end{center}    
	\vspace{0mm}
\end{table}
%
%
We corroborate our analysis on constrained maximization of achievable rate lower bounds with Matlab simulations. Our simulation parameters are given in Table \ref{table2}. We start by illustrating the  the behavior of our proposed power allocation schemes versus $\widehat{\nu}^*$. Fig. \ref{Pv_vs_v} shows the optimized $P$ obtained by solving \eqref{Rond_Lag_P} and the two proposed suboptimal schemes $P_{\rm{S}_2}$ and $P_{\rm{S}_1}$ versus $\widehat{\nu}^*$. We observe that $P_{\rm{S}_2}$ and $P_{\rm{S}_1}$  mimic the behavior of the optimized $P$. Furthermore, for the cut-off thresholds we have  $\zeta < \zeta_1 < \zeta_2$.
\par
\begin{figure}[!t]
	\centering
	\includegraphics[width=55mm]{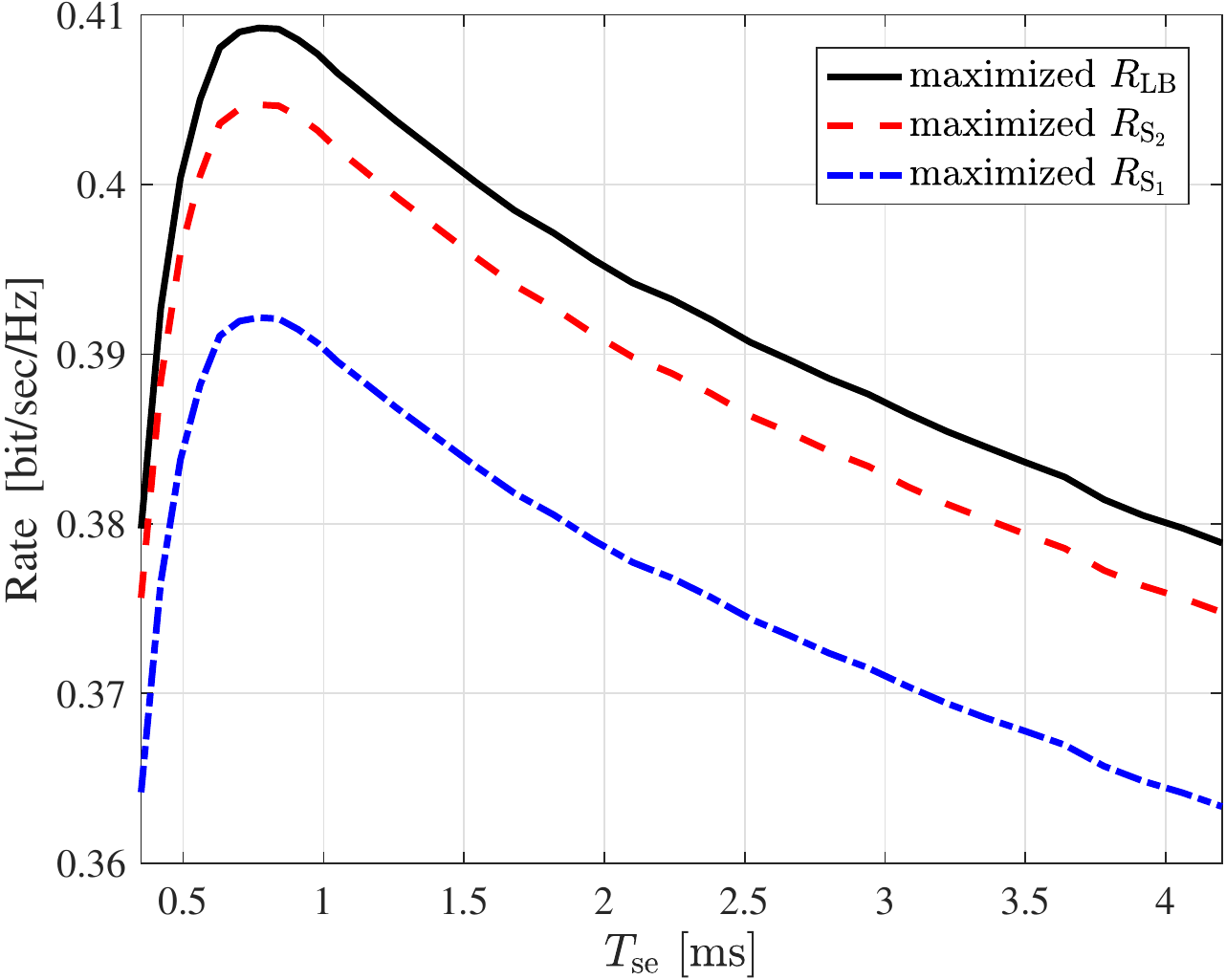}
	\caption{Rate versus $\Tsen$ for $\Pbar=2\,\text{dB}, \Ibar=-15\,$dB.} 
	\label{R_vs_Tsen}    
\end{figure}

 Next, we explore the effect of spatial spectrum sensing duration $\Tsen$ on the achievable rate lower bounds of our system. Fig. \ref{R_vs_Tsen} shows the maximized $\RLB, \Rnew$ and $\Rora$ (which we refer in the figures  to as ``Rate'') versus $\Tsen$. 
To plot this figure, we maximize the bounds w.r.t. only $\Ttra$ and $P$, subject to ATPC and AIC. We note that for all $\Tsen$ values we have $\RLB > \Rnew > \Rora$. We observe that the achievable rates always have a maximum in the interval $(0, \Tf- \Ttra)$. For the simulation parameters in Table \ref{table2} the optimized $\Tsen = 0.75\,\text{ms} =2.5\% \,\Tf$. Also, scheme 2 yields a higher achievable rate than that of scheme 1, because its corresponding power $P_{\rm{S}_2}$ fits better to the optimized power $P$ obtained from solving \eqref{Rond_Lag_P}. The achievable rate $\Rnew$ is very close to $\RLB$ and we do not have a significant performance loss if we choose the simple transmit power control scheme in \eqref{PNEW}.
%
%
%
%
%
%
\par
\begin{figure}[!t]
	\centering
	\includegraphics[width=55mm]{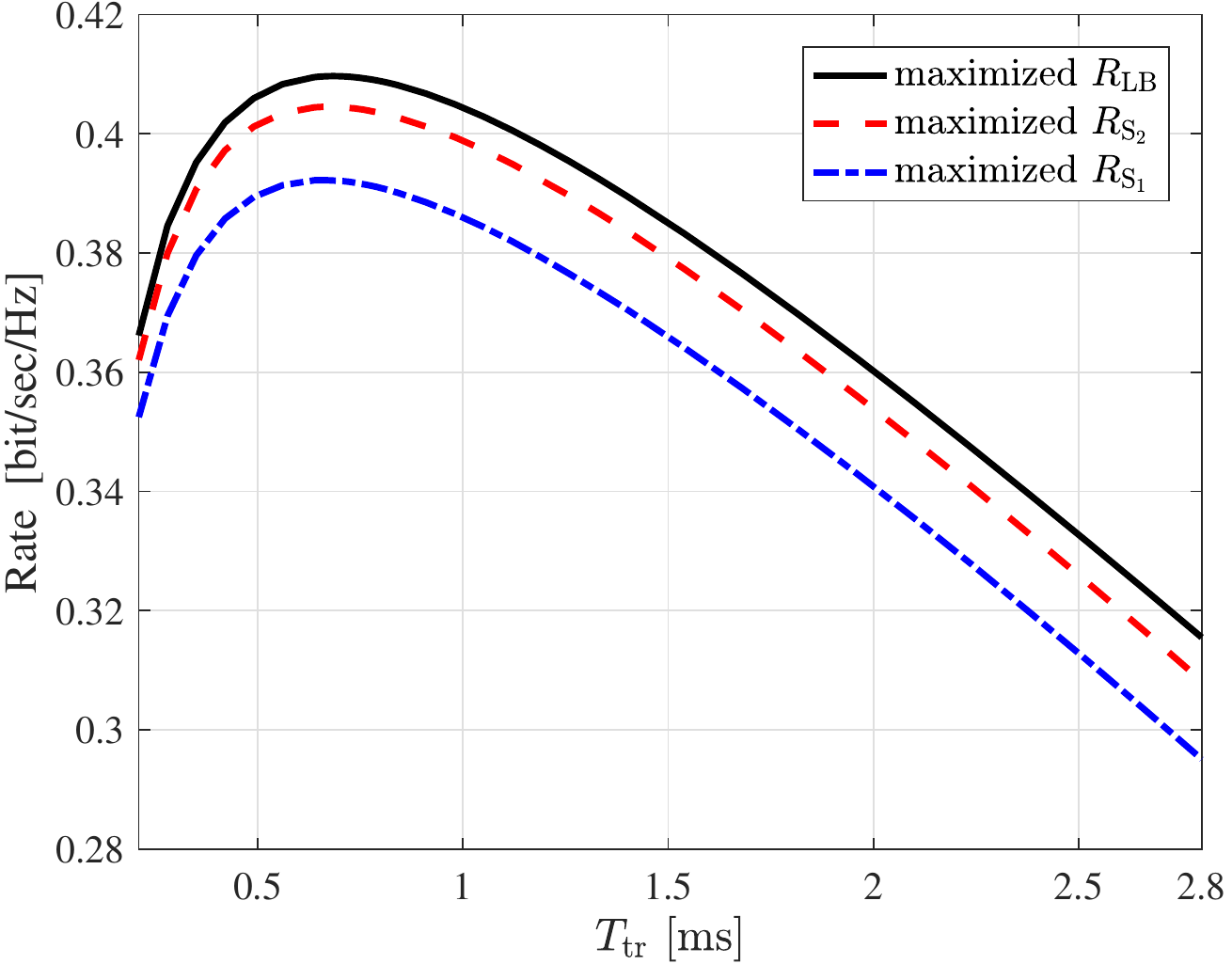}
	\caption{Rate versus $\Ttra$ for $\Pbar=2\,\text{dB}, \Ibar=-15\,$dB. } 
	\label{R_vs_Ttra}  
\end{figure}

 To investigate the effect of channel training duration $\Ttra$ on the achievable rate  lower bounds, we plot Fig. \ref{R_vs_Ttra} which illustrates the maximized $\RLB, \Rnew$ and $\Rora$ versus $\Ttra$.  To plot this figure, we maximize the bounds w.r.t. only $\Tsen$ and $P$, subject to ATPC and AIC. For all $\Ttra$ values we have $\RLB > \Rnew > \Rora$. We observe that the achievable rates always have a maximum in the interval $(0, \Tf- \Tsen)$.  For the simulation parameters in Table \ref{table2} the optimized $\Ttra = 0.67\,\text{ms} =2.23\% \,\Tf$. Comparing Fig. \ref{R_vs_Ttra} and Fig. \ref{R_vs_Tsen}, we notice that the achievable rates are more sensitive to the variations of $\Ttra$ compared to that of $\Tsen$. 
%
%
To be more specific, considering Fig. \ref{R_vs_Tsen} and Fig. \ref{R_vs_Ttra}, suppose we choose $\Tsen$ and $\Ttra$ values that are different from their corresponding maximum values by $20\%$, i.e., $\Delta \Tsen=20\%, \Delta \Ttra=20\%$. Then 
%
\begin{subequations}
\begin{align*}
\Delta \RLB/\Delta \Ttra > \Delta \RLB/\Delta \Tsen,\\
\Delta \Rnew/\Delta \Ttra > \Delta \Rnew/\Delta \Tsen,\\
\Delta \Rora/\Delta \Ttra > \Delta \Rora/\Delta \Tsen.
\end{align*}
\end{subequations}
%
These indicate that proper allocation of $\Ttra$ is more important than that of $\Tsen$, for providing higher achievable rates in our system. 
\par
\begin{figure}[!t] 
	\centering
	\includegraphics[width=55mm]{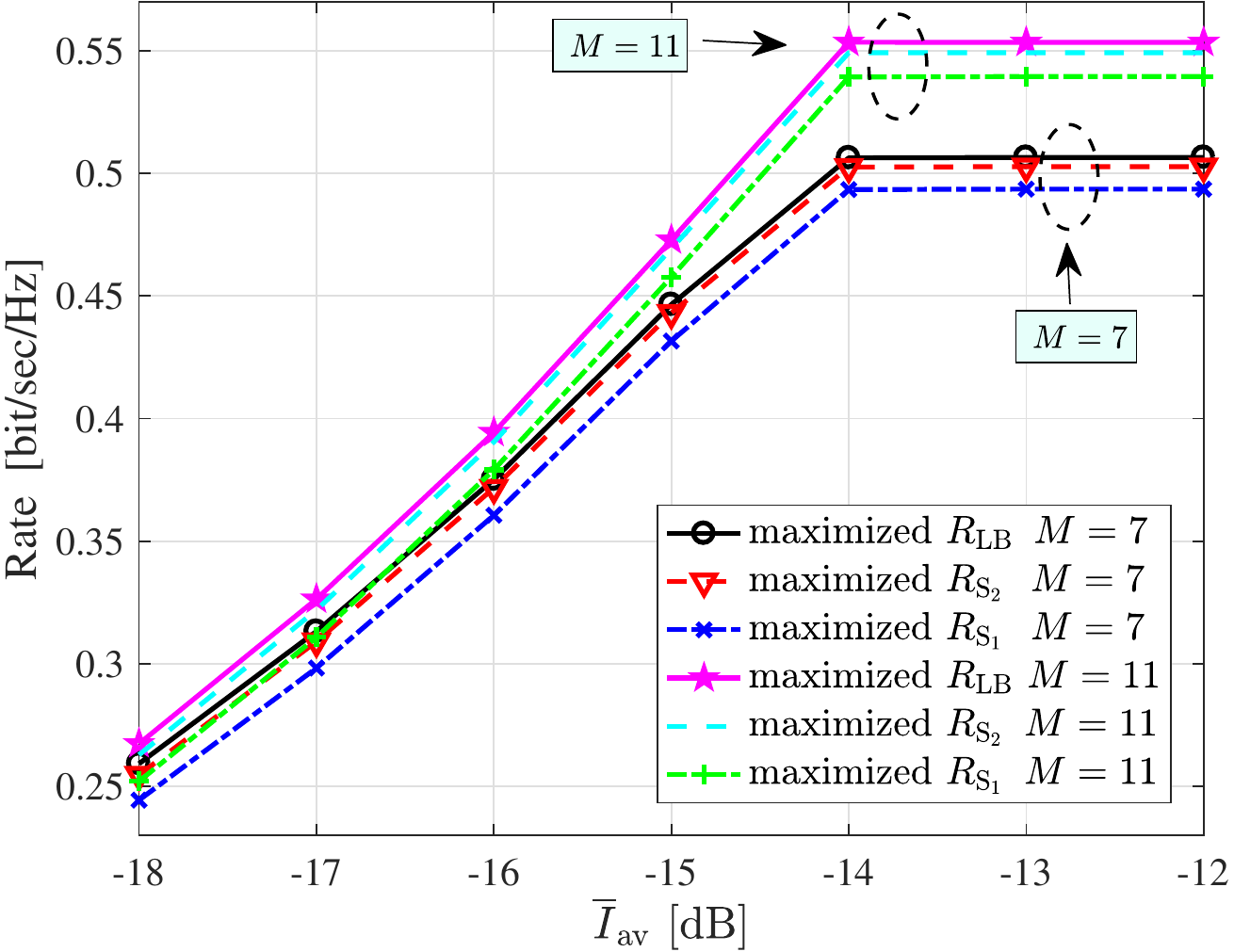}
	\caption{Rate versus $\Ibar$ for $M=7, 11$  and $\Pbar =2$\,dB.} 
	\label{R_vs_Iav}   
\end{figure}

\begin{figure}[!t] 
	\centering
	\includegraphics[width=55mm]{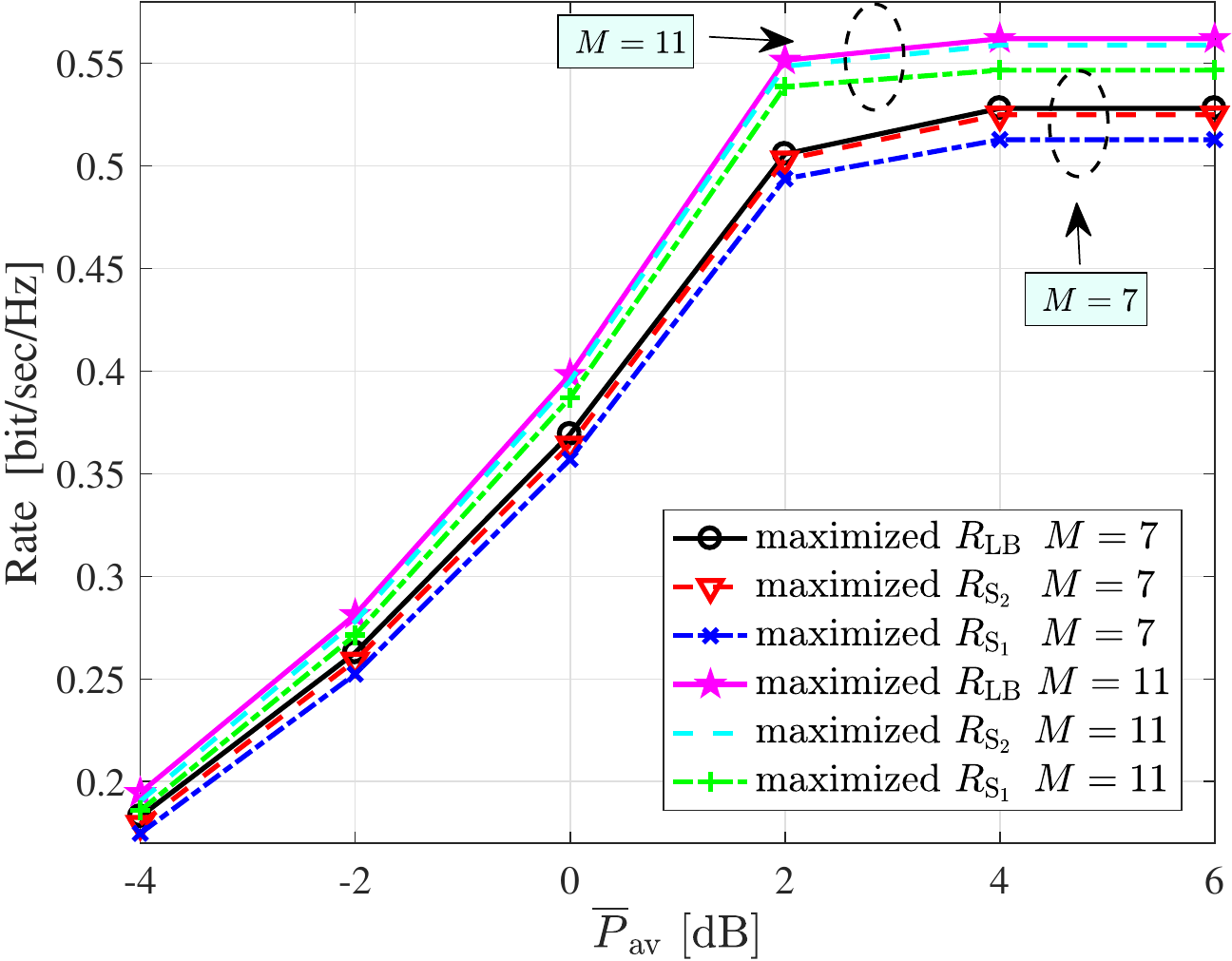}
	\caption{Rate versus $\Pbar$ for $M=7, 11$  and $ \Ibar = -14 $\,dB.} 
	\label{R_vs_Pav}   
\end{figure}

 To explore the effects of the number of beams $M$ and $\Ibar$ on the achievable rate  lower bounds, Fig. \ref{R_vs_Iav} illustrates the maximized $\RLB, \Rnew, \Rora$ versus $\Ibar$ for $M=7, 11$ and $\Pbar=2$\,dB. We observe that as $M$ increases a higher rate can be achieved. For all $M$ and $\Ibar$ values we have $\RLB > \Rnew > \Rora$. We realize that as $\Ibar$ increases from $-18\,$dB to $-14\,$dB, the achievable rates are monotonically increasing and the AIC is dominant. However, as $\Ibar$  increases beyond $-14\,$dB, the achievable rates remain unchanged and the ATPC is dominant. 
%
%
%
%
Fig. \ref{R_vs_Pav} illustrates the maximized $\RLB, \Rnew, \Rora$ versus $\Pbar$ for $M=7, 11$ and $\Ibar=-14$\,dB. The behaviors of the achievable rates in terms of $M$ are the same as Fig. \ref{R_vs_Iav}.  We note that as $\Pbar$ increases from $-4\,$dB to $2\,$dB, the achievable rates are monotonically increasing and the ATPC is dominant. However, as $\Pbar$  increases beyond $2\,$dB, the achievable rates remain unchanged and the AIC is dominant.
%
%
\par
\begin{figure}[!t]
	\centering
	\includegraphics[width=55mm]{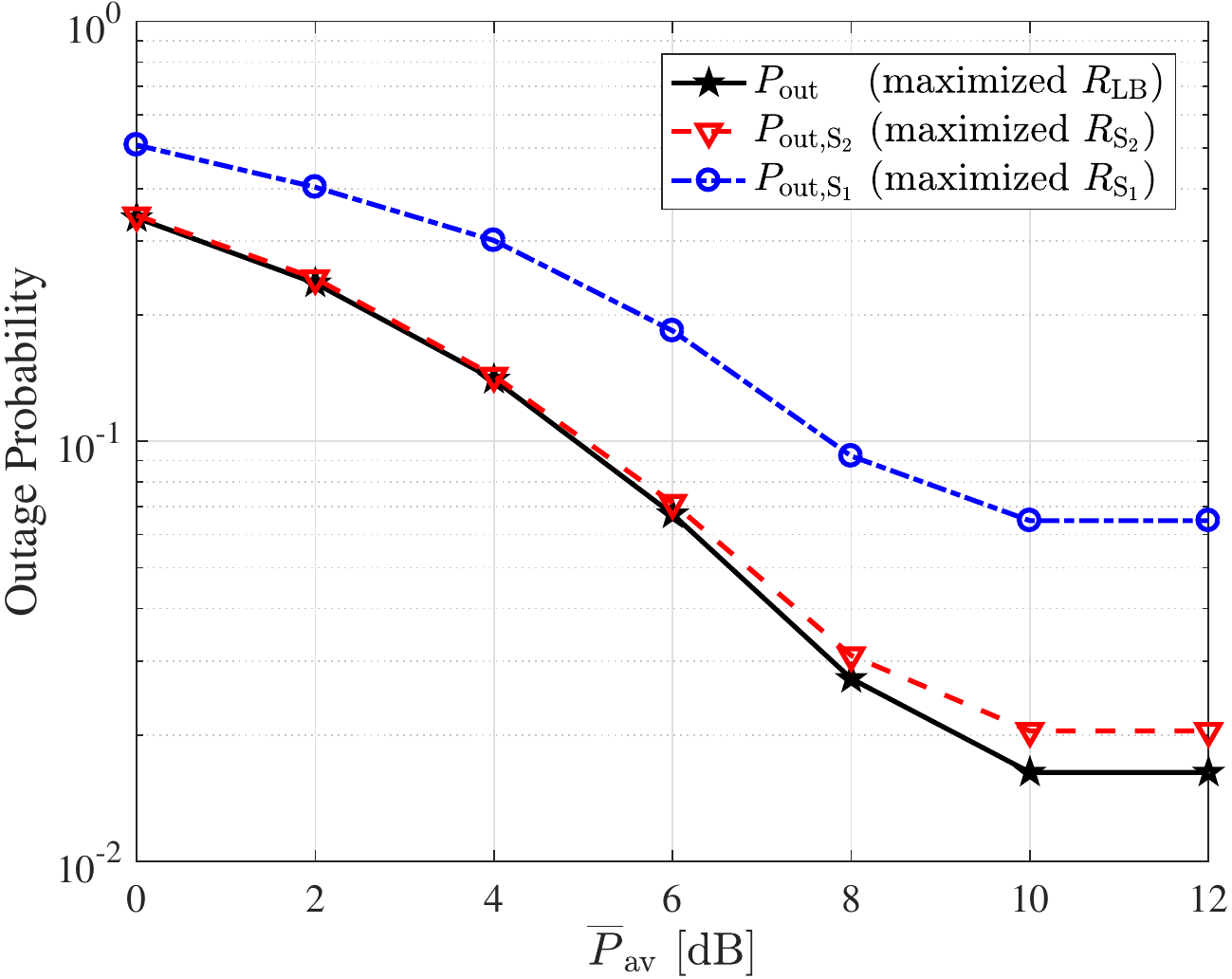}
	\caption{$P_{\rm{out}}$ versus $\Pbar$ for $\Ibar=-8\,$dB.} 
	\label{Pout_vs_Pav}    
\end{figure}

 We also consider outage probability as another performance metric to evaluate our system. We define the outage probability as the probability of \SUtx ~not transmitting data symbols due to the weak \SUtx-\SUrx ~channel when the spectrum is sensed idle, i.e., $P_{\rm{out}} = \Pr \big \{ P = 0 \,|\, \widehat{\cal H}_0 \big \}$. This probability can be directly obtained using the CDF of  $\widehat{\nu}^*$ evaluated at the cut-off threshold as the following
%
%
\begin{align*}
P_{\rm{out}} = & \Pr( \widehat{\nu}^* \leq \zeta | \widehat{\cal H}_0)= \omega_0 F^0_{ \widehat{\nu}^* } (\zeta) ~+ \omega_1 F^1_{ \widehat{\nu}^* } (\zeta), \\
P_{\rm{out},\rm{S}_1} = & \Pr( \widehat{\nu}^* \leq \zeta_1 | \widehat{\cal H}_0) =\omega_0 F^0_{ \widehat{\nu}^* } (\zeta_1) + \omega_1 F^1_{ \widehat{\nu}^* } (\zeta_1), \\
P_{\rm{out},\rm{S}_2} = & \Pr( \widehat{\nu}^* \leq \zeta_2 | \widehat{\cal H}_0) =\omega_0 F^0_{ \widehat{\nu}^* } (\zeta_2) + \omega_1 F^1_{ \widehat{\nu}^* } (\zeta_2). 
\end{align*}
%
%
\noindent Fig. \ref{Pout_vs_Pav}  illustrates $P_{\rm{out}}, P_{\rm{out},\rm{S}_2}, P_{\rm{out},\rm{S}_1}$ versus $\Pbar$ for $\Ibar=-8\,$dB. We observe that as $\Pbar$ increases the outage probabilities decrease. Moreover, for a given $\Pbar$ we have $P_{\rm{out}} < P_{\rm{out},\rm{S}_2} < P_{\rm{out},\rm{S}_1}$. This is consistent with Fig. \ref{Pv_vs_v} which shows for a given $\Pbar$, we have $\zeta < \zeta_1 < \zeta_2$. Combined this with the fact that the CFD $F_{ \widehat{\nu}^* } (\cdot)$  is  an increasing function of its argument, we reach the conclusion that $P_{\rm{out}} < P_{\rm{out},\rm{S}_2} < P_{\rm{out},\rm{S}_1}$.
%
%
%
%
\section{Conclusions}\label{Conclusion}

We considered an opportunistic CR system consisting of a PU, \SUtx, and \SUrx, where \SUtx ~is equipped with a RA that has $M$ beams, and there is an error-free low-rate feedback channel from \SUrx ~to \SUtx. We proposed a system design for integrated sector-based spatial spectrum sensing and sector-based data symbol communication.
We studied the entangled effects of spectrum sensing error, channel estimation error, and beam detection and beam selection errors (introduced by the RA), on the system achievable rates.  
We formulated a constrained optimization problem, where a lower bound on the achievable rate of \SUtx--\SUrx ~link is maximized, subject to ATPC and AIC, with the optimization variables being the durations of spatial spectrum sensing $\Tsen$ and  channel training $\Ttra$ as well as data symbol transmission power at \SUtx. 
Moreover, we proposed two alternative power adaptation schemes that are simpler to implement. We solved the proposed constrained optimization problems using iterative methods based on the BCD algorithm. Our simulation results demonstrate that one can increase  the achievable rates of \SUtx--\SUrx ~link significantly, via implementing these optimizations, while maintaining the ATPC and AIC.  They also showed that the achievable rates obtained from employing simple schemes 1 and 2 are very close to the one produced by the optimized transmit power. Our numerical results also showed that between optimizing $\Tsen$ and $\Ttra$, optimizing the latter has a larger effect on increasing the achievable rates in our system.
%
%
\section*{Acknowledgment}
This research was supported by NSF under grant ECCS-1443942.

\bibliographystyle{IEEEtran}
\bibliography{J2Ref}
%
\end{document}